\documentclass[useAMS,usenatbib]{mn2e}
\usepackage{graphicx}
\usepackage{aas_macros}
\usepackage{amssymb}

\newcommand{\about}{$\sim\!\!$~}

\newcommand{\be}{\begin{displaymath}}
\newcommand{\ee}{\end{displaymath}}

\def\lsim{\hbox{\rlap{\raise 0.425ex\hbox{$<$}}\lower 0.65ex\hbox{$\sim$}}}
\def\gsim{\hbox{\rlap{\raise 0.425ex\hbox{$>$}}\lower 0.65ex\hbox{$\sim$}}}
\def\arcmin{\hbox{$^\prime$}}

\newcommand{\msun}{M$_\odot$}

\newcommand{\etal}{et al.\ }

\newcommand{\kms}{km~s$^{-1}$}

\newcommand{\perMpc}{Mpc$^{-1}$}

\newcommand{\ion}[2]{#1$\;${\small{#2}}\relax}

\graphicspath{{/Users/jsilv/ptf13abc/plots/}}

\title[SN 2000cx and SN 2013bh: They Are Twins]{SN 2000cx and SN
  2013bh: Extremely Rare, Nearly Twin Type Ia Supernovae}

\author[Silverman, et al.]{Jeffrey M. Silverman,$^{1,2,3}$  Jozsef
  Vinko,$^{1,4}$ Mansi M. Kasliwal,$^{5}$ Ori D. Fox,$^{6}$ 
\newauthor
Yi Cao,$^{7}$  Joel Johansson,$^{8}$ Daniel A. Perley,$^{7}$ David Tal,$^{9}$ J. Craig Wheeler,$^{1}$ 
\newauthor 
Rahman Amanullah,$^{8}$ Iair Arcavi,$^{9}$ Joshua S. Bloom,$^{6}$ Avishay Gal-Yam,$^{9}$
\newauthor 
 Ariel Goobar,$^{8}$ Shrinivas R. Kulkarni,$^{7}$ Russ Laher,$^{10}$ William H. Lee,$^{11}$
\newauthor 
G. H. Marion,$^{1,12}$ Peter E. Nugent,$^{6,13}$ Isaac Shivvers$^{6}$ \\
$^{1}$Department of Astronomy, University of Texas at Austin, Austin, TX 78712, USA \\
$^{2}$NSF Astronomy and Astrophysics Postdoctoral Fellow \\
$^{3}$email: jsilverman@astro.as.utexas.edu \\
$^{4}$Department of Optics and Quantum Electronics, University of Szeged, D\'{o}m t\'{e}r 9, 6720 Szeged, Hungary \\
$^{5}$Observatories of the Carnegie Institution of Science, Pasadena, CA 91101, USA \\
$^{6}$Department of Astronomy, University of California, Berkeley, CA 94720-3411, USA \\
$^{7}$Cahill Center for Astrophysics, California Institute of Technology, Pasadena, CA 91125, USA \\
$^{8}$The Oskar Klein Centre, Department of Physics, AlbaNova, Stockholm University, SE-106 91 Stockholm, Sweden \\
$^{9}$Benoziyo Center for Astrophysics, The Weizmann Institute of Science, Rehovot 76100, Israel \\
$^{10}$Spitzer Science Center, California Institute of Technology, MC 314-6, Pasadena, CA 91125, USA \\
$^{11}$Instituto de Astronom\'ia, Universidad Nacional Aut\'onoma de M\'exico, Apartado Postal 70-264, 04510 M\'exico, D.F., M\'exico \\
$^{12}$Harvard-Smithsonian Center for Astrophysics, 60 Garden St., Cambridge, MA 02138, USA \\
$^{13}$Lawrence Berkeley National Laboratory, Berkeley, CA 94720, USA }

\begin{document}
\date{Accepted  . Received   ; in original form  }
\pagerange{\pageref{firstpage}--\pageref{lastpage}} \pubyear{2013}
\maketitle
\label{firstpage}

\begin{abstract}
The Type~Ia supernova (SN~Ia) SN~2000cx was one of the most peculiar
transients ever discovered, with a rise to maximum brightness typical
of a SN~Ia, but a slower decline and a higher photospheric
temperature. Thirteen years later SN~2013bh (aka iPTF13abc), a near
identical twin, was discovered and we obtained optical and near-IR
photometry and low-resolution optical spectroscopy from discovery
until about 1 month past $r$-band maximum brightness. The spectra of
both objects show iron-group elements (\ion{Co}{II}, \ion{Ni}{II},
\ion{Fe}{II}, \ion{Fe}{III}, and high-velocity features [HVFs] of
\ion{Ti}{II}), intermediate-mass elements (\ion{Si}{II},
\ion{Si}{III}, and \ion{S}{II}), and separate normal velocity features
(\about12000~\kms) and HVFs (\about24000~\kms) of
\ion{Ca}{II}. Persistent absorption from \ion{Fe}{III} and
\ion{Si}{III}, along with the colour evolution, imply high blackbody
temperatures for SNe~2013bh and 2000cx (\about12000~K). Both objects
lack narrow \ion{Na}{I}~D absorption and exploded in the outskirts of
their hosts, indicating that the SN environments were relatively free
of interstellar or circumstellar material and may imply that the
progenitors came from a relatively old and low-metallicity 
stellar population. Models of SN~2000cx, seemingly applicable to
SN~2013bh, imply the production of up to \about1~\msun\ of $^{56}$Ni
and 
(4.3--5.5)$\times 10^{-3}$~\msun\ of fast-moving Ca ejecta. 
\end{abstract}




\section{Introduction}\label{s:intro}

Resulting from the thermonuclear explosion of C/O white dwarfs
\citep[WDs; e.g.,][]{Nugent11,Bloom12}, 
Type~Ia supernovae (SNe~Ia) provided the first clear indication that
the expansion of the Universe is accelerating
\citep{Riess98:lambda,Perlmutter99} and have been used as precise
distance indicators to accurately measure cosmological parameters
\citep[e.g.,][]{Conley11,Sullivan11,Suzuki12}. Despite their utility,
the specifics of SN~Ia progenitor systems and explosion mechanisms are
still unclear \citep[see][for further information]{Howell11}. In
general, the two leading progenitor scenarios are the
single-degenerate (SD) channel, when the WD accretes matter from a
non-degenerate companion star \citep[e.g.,][]{Whelan73}, and the
double-degenerate (DD) channel, which is the result of the merger of
two WDs \citep[e.g.,][]{Iben84,Webbink84}.

The ability to determine cosmological distances using SNe~Ia lies
mainly in the fact that they follow a light-curve decline rate
versus peak luminosity correlation \citep[i.e., the ``Phillips
relation'';][]{Phillips93}; however, the scatter in this relation is
at least partially caused by the inclusion of various peculiar SNe~Ia
that nominally follow the correlation.  
Some of these objects, which are definitively classified as SNe~Ia,
can show extreme peculiarities. One of the most well-known and
infamous of these objects is SN~2000cx.

SN~2000cx was discovered using the 0.76 m Katzman Automatic Imaging
Telescope (KAIT; \citealt{Richmond93}) on 2000~Jul.~17.5 \citep[][UT
dates are used throughout]{00cx:disc}, as part of the Lick Observatory
Supernova Search (LOSS; \citealt{Filippenko01}). It occurred in the
outskirts of the nearby ($z = 0.00807$) S0 galaxy NGC~524 and was
intensely observed by multiple groups
\citep[e.g.][]{Li01:00cx,Candia03}. It was soon discovered that
SN~2000cx was peculiar in many ways. Its rise to maximum brightness in
the $B$ band was typical for a SN~Ia, but it declined much slower than
normal. Thus standard light curve fitting algorithms could not be
reliably used on SN~2000cx \citep{Li01:00cx}. Furthermore, the optical 
colours of the SN were redder than nearly all other SNe~Ia near maximum
brightness, but quickly became bluer than other SNe~Ia by 15~d after
maximum  \citep{Candia03}.

Spectroscopically, SN~2000cx showed strong \ion{Fe}{III} and weak
\ion{Si}{II} and \ion{S}{II} features, and thus resembled the overluminous
SN~1991T \citep{Filippenko92:91T,Phillips92}. Unlike SN~1991T (and
pretty much all other SNe~Ia ever observed), SN~2000cx continued to
show strong \ion{Fe}{III} features (indicative of a high photospheric
temperature) through 20~d past maximum brightness
\citep{Li01:00cx}. In addition, high-velocity features (HVFs) of
\ion{Ca}{II} were observed at velocities $> 20000$~\kms, separated from
a second set of \ion{Ca}{II} features at more typical velocities
\citep[\about12000~\kms,][]{Thomas04,Branch04:00cx}, the latter of
which we will refer to as the photospheric velocity features
(PVFs). These extreme observables indicate that SN~2000cx likely
produced $> 0.5$~\msun\ of $^{56}$Ni \citep{Sollerman04} and possibly
up to \about1~\msun\ of $^{56}$Ni \citep{Li01:00cx}. SN~2000cx
remained unique--until now.

Herein we present optical and near-IR photometry and low-resolution
optical spectra of a near twin of SN~2000cx, SN~2013bh (aka
iPTF13abc), during the first month after its discovery. We discuss 
the discovery of this object in \S\ref{s:discovery} and present our
observations and describe our data reduction in
\S\ref{s:observations}. Our analysis of the data, as well as a
discussion of the host galaxy and possible progenitor system of
SN~2013bh, can be found in \S\ref{s:analysis}. We summarize our
conclusions in \S\ref{s:conclusions}.

\begin{figure*}
\centering
\includegraphics[width=3in]{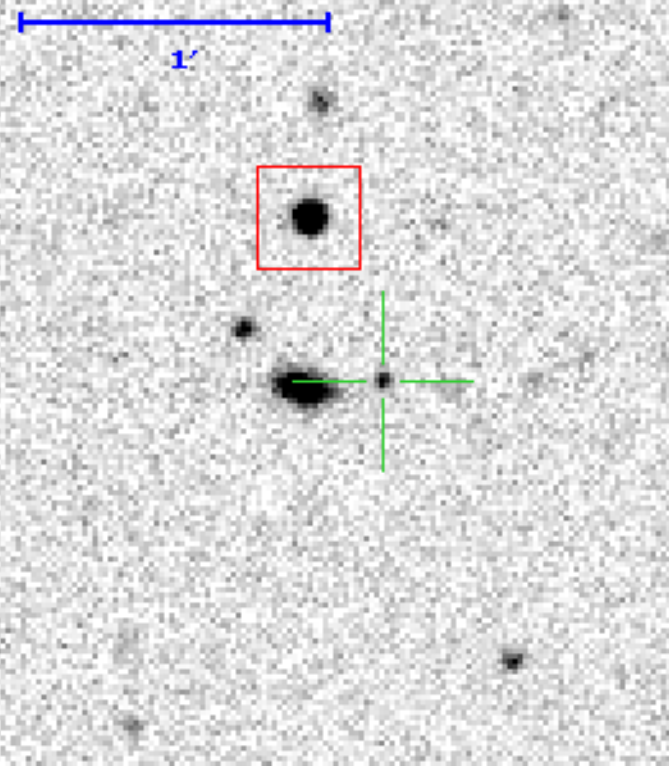}
\caption{The field of iPTF13abc with the SN marked with the green
  cross-hairs and a bright offset star marked with the red square. Up
  is north and left is east. The host galaxy of iPTF13abc is east and
  slightly south of the SN.}\label{f:field} 
\end{figure*}

\section{Discovery}\label{s:discovery}

iPTF13abc was discovered on 2013~Mar.~23.5 with $r = 19.64 \pm
0.14$~mag as part of the intermediate Palomar Transient Factory
\citep[iPTF;][]{Law09,Rau09,Kulkarni13}. It was found at J2000.0 
coordinates $\alpha = 15^{\mathrm{h}} 02^{\mathrm{m}} 13.09^{\mathrm{s}}$,
$\delta = +10^{\circ} 38\arcmin 45\farcs6$ in the star-forming galaxy 
SDSS~J150214.17+103843.6 \citep[$z = 0.0744$,][]{Ahn12}. The field of
iPTF13abc is shown in 
Figure~\ref{f:field} with the SN marked with the green cross-hairs and
a bright offset star marked with the red square. Up is north and left
is east; the host galaxy of iPTF13abc is east and slightly south of
the SN.

 About 10 days later, this   
SN was independently discovered by the Catalina Real-Time Transient
Survey \citep[CRTS;][]{Drake09} as CSS130403:150213+103846, and upon
public announcement of the discovery was christened SN~2013bh 
\citep{13bh:disc}. On 2013~Apr.~1.3, we obtained a spectrum of
SN~2013bh and found that it was similar to the aforementioned
SN~2000cx a few days before $R$-band maximum brightness. Three days
later, the Public ESO Spectroscopic Survey of Transient Objects
\citep[PESSTO;
e.g.,][]{Maund12}\footnote{http://www.pessto.org/pessto/index.py} 
collaboration obtained a spectrum of this object and also found
SN~2013bh to be spectroscopically similar to SN~2000cx
\citep{13bh:disc2}.


\section{Observations and Data Reduction}\label{s:observations}

\subsection{Photometry}\label{ss:phot}

SN~2013bh was discovered using the 48~in Samuel Oschin Telescope at
Palomar Observatory (P48). Every night since 15~May~2013, P48 observed 
the field of SN~2013bh three times in the $R$-band,\footnote{All
  photometric data presented herein are in AB mags, except the $J$-
  and $H$-band data which are in Vega mags.}
weather-permitting. Adjacent visits were separated by at least 
45~min. The P48 images of SN~2013bh were processed with the iPTF LBL
pipeline (Nugent \etal in preparation). Aperture photometry was then
performed on each of these images with an aperture radius equal to the
seeing of each image. Aperture correction coefficients were calculated
by measuring unsaturated stars that have signal-to-noise ratio (S/N)
$> 20$ with two apertures, one of which had a radius of seeing while 
the other had a radius of three times seeing. Relative photometry was
done among all P48 images while the absolute photometry was calibrated
to $r$-band data of the Sloan Digital Sky Survey (SDSS) Data Release 9 
(DR9) Catalogue \citep{Ahn12}. The P48 $R$-band data of SN~2013bh is
presented in Table~\ref{t:p48}.

\begin{table}
\begin{center}
\caption{P48 Photometry of SN~2013bh\label{t:p48}}
\begin{tabular}{lc}
\hline\hline
JD-2456000 & $R$ (mag) \\
\hline
366.88 & $<$21.22 \\
370.02 & $<$21.46 \\
375.02 & 19.70 (0.14) \\
376.03 & 19.43 (0.10) \\
377.01 & 19.22 (0.11) \\
386.98 & 18.35 (0.24) \\
392.95 & 18.49 (0.11) \\
393.78 & 18.50 (0.11) \\
396.94 & 18.80 (0.13) \\
\hline \hline
\multicolumn{2}{l}{1$\sigma$ uncertainties are in parentheses.} \\
\hline\hline
\end{tabular}
\end{center}
\end{table}

After discovery, multiple telescopes were employed in order to follow
SN~2013bh photometrically. As part of regular iPTF operations
\citep{Law09}, robotic multi-band (\protect\hbox{$g\!ri$}) follow-up
photometry was obtained with the Palomar 60~in (P60). First, the
images were reduced using the automated P60 pipeline \citep{Cenko06}.
Next, calibration star magnitudes were measured on the best seeing
image. Finally, aperture photometry on SN~2013bh was undertaken (with
the radius equal to the seeing and the background annulus equal to 3
and 5 times the seeing) and calibrated relative to these stars. Given
the offset from the host galaxy, image subtraction was not
required. Table~\ref{t:p60} shows our P60 photometry of SN~2013bh. 

\begin{table}
\begin{center}
\caption{P60 Photometry of SN~2013bh\label{t:p60}}
\begin{tabular}{lccc}
\hline\hline
JD-2456000 & $g$ (mag) & $r$ (mag) & $i$ (mag) \\
\hline
375.40 & $\cdots$ & 19.42 (0.05) & $\cdots$ \\
388.44 & 18.38 (0.02) & 18.31 (0.03) & 18.94 (0.06) \\
389.34 & 18.41 (0.03) & 18.31 (0.03) & 18.93 (0.05) \\
390.25 & 18.44 (0.05) & 18.34 (0.04) & 19.14 (0.11) \\
392.21 & 18.53 (0.06) & 18.36 (0.05) & 19.10 (0.12) \\
393.21 & $\cdots$ & 18.46 (0.02) & 19.08 (0.05) \\
406.17 & $\cdots$ & 19.39 (0.14) & $\cdots$ \\
409.19 & 20.11 (0.42) & 19.33 (0.11) & $\cdots$ \\
412.42 & 20.18 (0.08) & 19.50 (0.05) & 19.89 (0.09) \\
413.41 & $\cdots$ & $\cdots$ & 19.85 (0.07) \\
414.31 & 20.32 (0.06) & $\cdots$ & $\cdots$ \\
415.32 & $\cdots$ & $\cdots$ & 19.91 (0.08) \\
416.22 & 20.54 (0.06) & 19.60 (0.08) & $\cdots$ \\
417.36 & $\cdots$ & $\cdots$ & 19.95 (0.15) \\
422.38 & 21.06 (0.08) & 20.17 (0.06) & 20.39 (0.11) \\
423.36 & 21.11 (0.10) & 20.20 (0.06) & 20.38 (0.10) \\
426.22 & 21.13 (0.10) & 20.32 (0.06) & 20.51 (0.11) \\
427.21 & 21.26 (0.15) & 20.43 (0.08) & 20.62 (0.15) \\
\hline\hline
\multicolumn{4}{l}{1$\sigma$ uncertainties are in parentheses.} \\
\hline\hline
\end{tabular}
\end{center}
\end{table}

Using ALFOSC at the Nordic Optical Telescope (NOT), La Palma, we
obtained \protect\hbox{$BV\!RI$} photometry of SN~2013bh. All data
have been reduced with standard 
IRAF\footnote{IRAF: The Image Reduction and Analysis Facility is
  distributed by the National Optical Astronomy Observatory, which is
  operated by the Association of Universities for Research in
  Astronomy (AURA) under cooperative agreement with the National
  Science Foundation (NSF).} routines, using the QUBA pipeline
\citep[see][for more information]{Valenti11}. The magnitudes are
measured with aperture photometry (with the radius equal to the 2
times the seeing and the background annulus equal to 3 and 5 times the
seeing) and calibrated to the Landolt system through observations of
standard stars PG1047+003 and PG1525-071. In order to match the other
SN~2013bh photometry, the $RI$ data from NOT were converted to $ri$
magnitudes using the conversions presented by \citet{Jordi06}. These
data are presented in Table~\ref{t:not}.

\begin{table*}
\begin{center}
\caption{NOT Photometry of SN~2013bh\label{t:not}}
\begin{tabular}{lcccc}
\hline\hline
JD-2456000 & $B$ (mag) & $V$ (mag) & $r$ (mag) & $i$ (mag) \\
\hline
396.65 & 18.98 (0.03) & 18.48 (0.03) & 18.78 (0.03) & 19.77 (0.03) \\
407.45 & 19.69 (0.09) & 19.27 (0.09) & $\cdots$ & $\cdots$ \\
\hline\hline
\multicolumn{5}{p{5in}}{1$\sigma$ uncertainties are in parentheses. $ri$
  magnitudes have been converted from $RI$ data using the conversions
  presented by \citet{Jordi06}.} \\
\hline\hline
\end{tabular}
\end{center}
\end{table*}

Optical and near-IR photometry (\protect\hbox{$riZY\!JH$}) were
obtained with the multi-channel Reionization And Transients InfraRed
camera \citep[RATIR;][]{Butler12} mounted on the 1.5~m Johnson
telescope at the Mexican Observatorio Astrono\'mico Nacional on Sierra
San Pedro M\'artir in Baja California, M\'exico
\citep{Watson12}. Typical observations include a series of 60~s
exposures in \protect\hbox{$riZY\!JH$}, with dithering between
exposures. The fixed IR filters of RATIR cover half of their
respective detectors, automatically providing off-target IR sky
exposures while the target is observed in the neighboring
filter. Master IR sky frames are created from a median stack of
off-target images in each IR filter. No off-target sky frames were
obtained on the optical CCDs, but the small galaxy size and sufficient
dithering allowed for a sky frame to be created from a median stack of
all the images in each filter. Flat-field frames consist of evening
sky exposures. Given the lack of a cold shutter in the RATIR design,
IR darks are not available. Laboratory testing, however, confirms that
dark current is negligible in both IR detectors \citep{Fox12}.

The RATIR data were reduced, co-added, and analysed using standard CCD
and IR processing techniques in IDL and Python, utilizing online
astrometry programs {\tt SExtractor} and {\tt
  SWarp}\footnote{SExtractor and SWarp can be accessed from
  http://www.astromatic.net/software .}. Calibration was performed
using field stars with reported fluxes in both 2MASS
\citep{Skrutskie06} and the SDSS DR9 Catalogue
\citep{Ahn12}. Table~\ref{t:ratir} lists our RATIR photometry of
SN~2013bh.

\begin{table*}
\begin{center}
\caption{RATIR Photometry of SN~2013bh\label{t:ratir}}
\begin{tabular}{lcccccc}
\hline\hline
JD-2456000 & $r$ (mag) & $i$ (mag) & $Z$ (mag) & $Y$ (mag) & $J$ (mag) & $H$ (mag) \\
\hline
386.00 & 18.34 (0.01) & 18.88 (0.01) & $\cdots$ & 19.15 (0.05) & $\cdots$ & $<$18.50 \\
389.00 & 18.26 (0.01) & 18.93 (0.01) & $\cdots$ & 19.39 (0.05) & $\cdots$ & $<$18.50 \\
397.00 & 18.61 (0.03) & 19.66 (0.03) & $\cdots$ & $\cdots$ & $\cdots$ & $\cdots$ \\
398.00 & 18.64 (0.01) & 19.56 (0.02) & $\cdots$ & 19.95 (0.11) & $\cdots$ & $<$18.50 \\
399.00 & 18.77 (0.01) & 19.73 (0.02) & 19.98 (0.07) & 19.93 (0.12) & $<$19.40 & $<$18.50 \\
410.00 & $\cdots$ & 19.88 (0.02) & 20.45 (0.10) & 19.73 (0.06) & $<$19.90 & $<$18.50 \\
423.00 & 20.30 (0.02) & 20.25 (0.03) & 20.65 (0.12) & 20.02 (0.11) & $<$19.30 & $<$18.50 \\
\hline\hline
\multicolumn{7}{l}{1$\sigma$ uncertainties are in parentheses.} \\
\hline\hline
\end{tabular}
\end{center}
\end{table*}

In addition to obtaining our own data described above, we 
also downloaded publicly available $V$-band data from CRTS 
(Table~\ref{t:crts}).\footnote{http://nesssi.cacr.caltech.edu/catalina/20130403/1304031090804140600.html} 

\begin{table}
\begin{center}
\caption{CRTS Photometry of SN~2013bh\label{t:crts}}
\begin{tabular}{lc}
\hline\hline
JD-2456000 & $V$ (mag) \\
\hline
385.50 & 18.44 (0.07) \\
392.50 & 18.48 (0.17) \\
402.50 & 19.16 (0.96) \\
411.50 & 19.19 (0.77) \\
\hline\hline
\multicolumn{2}{l}{1$\sigma$ uncertainties are in parentheses.} \\
\hline\hline
\end{tabular}
\end{center}
\end{table}

\subsection{Spectroscopy}\label{ss:spec}

A spectroscopic time series of SN~2013bh was also obtained soon after
discovery. Low-resolution optical spectra were obtained mainly using
the Marcario Low-Resolution Spectrograph \citep[LRS;][]{Hill98} on the
9.2~m Hobby-Eberly Telescope (HET) at McDonald Observatory, though
spectral data were also acquired with the Low Resolution Imaging
Spectrometer \citep[LRIS;][]{Oke95} on the 10~m Keck~I telescope and
the Double Beam Spectrograph \citep[DBSP;][]{Oke82} on the Palomar
200~in telescope. We also downloaded, via WISeREP \citep[the Weizmann 
Interactive Supernova data
REPository;][]{Yaron12},\footnote{http://www.weizmann.ac.il/astrophysics/wiserep}
a publicly available spectrum of SN~2013bh obtained by the PESSTO
collaboration using the ESO Faint Object Spectrograph and Camera v.2
\citep[EFOSC2;][]{Buzzoni84} on the ESO New Technology Telescope
(NTT). Table~\ref{t:spec} summarizes the spectral data of SN~2013bh
presented here and upon publication all of these spectra will be
available in electronic format on WISeREP.

\begin{table*}
\begin{center}
\caption{Journal of Spectroscopic Observations of SN~2013bh}\label{t:spec}
\begin{tabular}{lrlcrr}
\hline \hline
UT Date &  Epoch$^\textrm{a}$  &  Instrument & Range (\AA)  & Res. (\AA)$^\textrm{b}$ &  Exposure (s) \\
\hline
2013~Apr.~1.31 & $-3.6$ & LRS & 4192--10204 & 15.6 & 2000 \\
2013~Apr.~5.40 & 0.2 & EFOSC2 & 3764--9283$\phantom{1}$ & 18 & 1500 \\
2013~Apr.~6.46 & 1.2 & LRS & 4192--10128 & 15.4 & 2000 \\
2013~Apr.~9.56 & 4.1 & LRIS & 3764--10204 & 4.5/6 & 480 \\
2013~Apr.~11.44 &	5.8 & LRS & 4192--10204 & 15.5 & 2000 \\
2013~Apr.~13.48 &	7.7 & DBSP & 3624--9872$\phantom{1}$ & 3/4 & 300 \\
2013~Apr.~16.43 &	10.5 & LRS & 4192--10204 & 15.2 & 2000 \\
2013~Apr.~22.26 &	15.9 & LRS & 4204--10204 & 15.6	& 2000 \\
2013~Apr.~27.23 & 20.5 & LRS & 4288--10200 & 15.9 & 2500 \\
\hline\hline
\multicolumn{6}{p{5.6in}}{$^\textrm{a}$Rest-frame days relative to the
  date of $r$-band maximum (2013~Apr.~5.2).} \\
\multicolumn{6}{p{5.6in}}{$^\textrm{b}$LRS = Low-Resolution
  Spectrograph on the 9.2~m Hobby-Eberly Telescope at McDonald
  Observatory; EFOSC2 = ESO Faint Object Spectrograph and Camera v.2
  on the NTT; LRIS = Low Resolution Imaging Spectrometer on the 10~m
  Keck~I telescope; DBSP = Double Spectrograph on the Palomar 200~in
  telescope.} \\ 
\multicolumn{6}{p{5.6in}}{$^\textrm{b}$Approximate full width at
  half-maximum intensity (FWHM) resolution. If two numbers are listed,
  they represent the blue- and red-side resolutions, respectively.} \\
\hline\hline
\end{tabular}
\end{center}
\end{table*}

All of our spectra were reduced using standard techniques
\citep[e.g.,][]{Silverman12:BSNIPI}. Routine CCD processing and
spectrum extraction were completed with IRAF. We obtained the
wavelength scale from low-order polynomial fits to calibration-lamp
spectra. Small wavelength shifts were then applied to the data after
cross-correlating a template sky to the night-sky lines that were
extracted with the SN. Using our own IDL routines, we fit
spectrophotometric standard-star spectra to the data in order to flux
calibrate our spectra and to remove telluric lines
\citep{Wade88,Matheson00:93j}.


\section{Analysis and Results}\label{s:analysis}

\subsection{Light Curves}\label{ss:lightcurves}

\begin{figure*}
\centering
\includegraphics[width=6in]{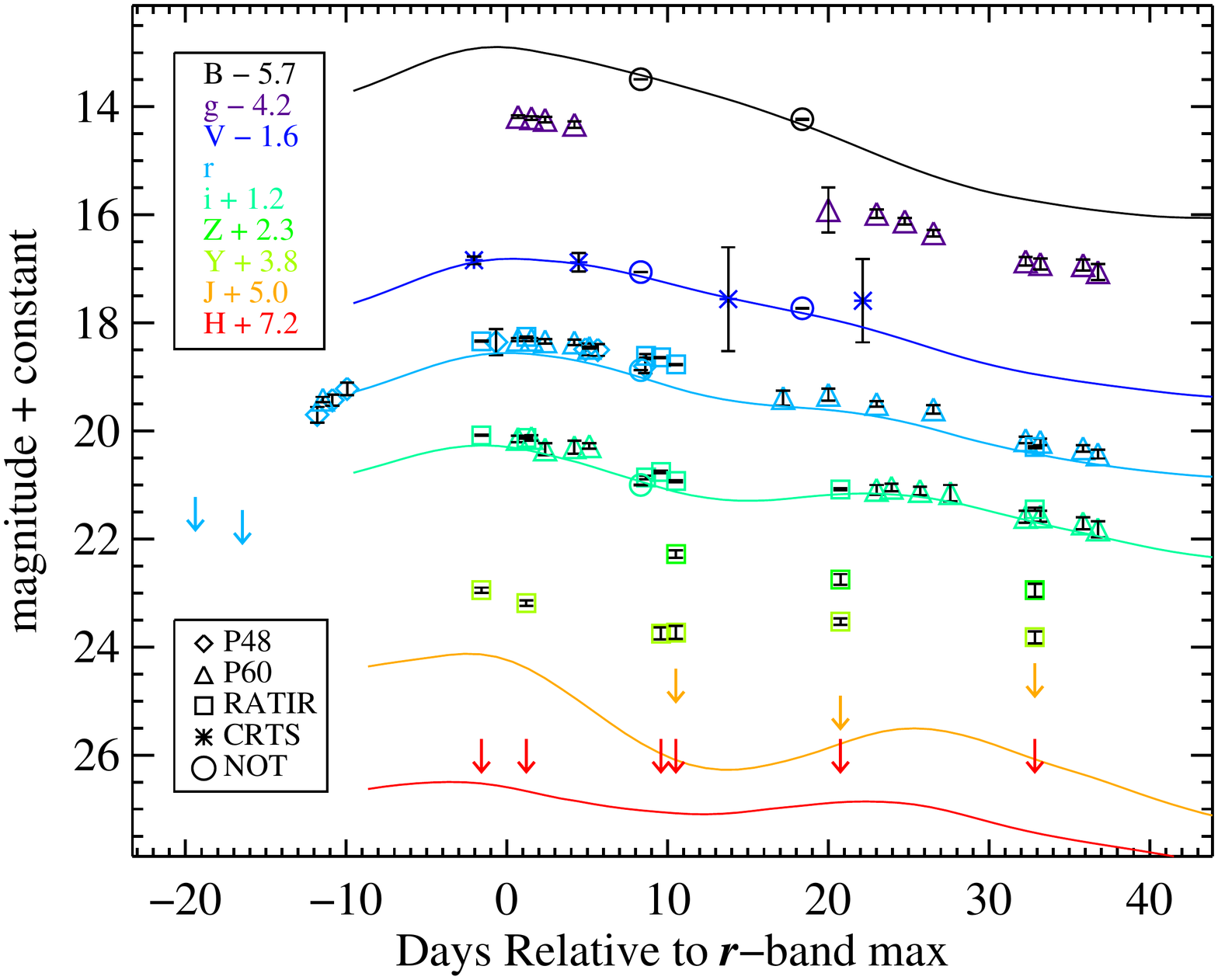}
\caption{\protect\hbox{$BgV\!riZY\!JH$} light curves of SN~2013bh
  (points), compared with those of SN~2000cx \citep[solid
  lines,][]{Li01:00cx,Candia03}. The SN~2000cx data has been shifted
  in time to match the $r$-band peak of SN~2013bh and shifted in
  apparent magnitude to the distance of SN~2013bh. Neither of the
  objects have been dereddened.}\label{f:lcs} 
\end{figure*}

We present our \protect\hbox{$BgV\!riZY\!JH$} light curves of SN~2013bh
in Figure~\ref{f:lcs} (points), compared with those of SN~2000cx
\citep[solid lines,][]{Li01:00cx,Candia03}. Upper limits are shown as
downward-pointing arrows. In the figure, the SN~2000cx data has been
shifted in time to match the $r$-band peak of SN~2013bh. It has
also been shifted in apparent magnitude to the distance of SN~2013bh,
however no other vertical shift has been applied. Neither of the
objects have been dereddened. The photometric data of SN~2000cx
matches SN~2013bh in all bands for which there are observations of
both objects. The observed brightness of the two objects is consistent
in $B$ and $V$, while SN~2013bh appears to be $\la 0.3$~mag brighter
than SN~2000cx in $r$ and $i$. The upper limits on SN~2013bh in the
$J$ and $H$ bands are also consistent with the observed magnitudes of
SN~2000cx in these bandpasses.

From low-order polynomial fits, we find that SN~2013bh reached a peak
$r$-band magnitude of $18.3 \pm 0.03$ on 2013~Apr.~$5.2 \pm 0.5$ and a
peak $i$-band magnitude of $18.9 \pm 0.3$ on 2013~Apr.~$4.5 \pm
1$. The peak values for the $i$ band are somewhat uncertain due to the
fact that our earliest data for this bandpass are very close to maximum
brightness. SN light-curve declines are often described by $\Delta
m_{15}$, i.e., the decrease in magnitudes in a given bandpass from
peak to 15~d past peak. SN~2013bh has $\Delta m_{15}(r) = 0.73 \pm
0.03$~mag and $\Delta m_{15}(i) = 0.96 \pm 0.10$~mag, which are both
somewhat smaller than the corresponding values for SN~2000cx
\citep[0.94 and 1.06~mag, respectively;][]{Li01:00cx}, implying that
the light curve of SN~2013bh is slightly broader than that of
SN~2000cx.

Given $E(B-V)_\textrm{MW} = 0.0284$~mag \citep{Schlegel98}, no
reddening from the host galaxy of SN~2013bh (see
Section~\ref{ss:spectra}), and $\mu = 37.57 \pm 0.15$~mag (see
Section~\ref{ss:host}), the peak absolute magnitudes of SN~2013bh in
our best observed bands are $M_r = -19.3 \pm 0.15$~mag and $M_i =
-18.7 \pm 0.34$~mag. These values are in agreement with those of
SN~2000cx, after correcting for $E(B-V)_\textrm{MW} = 0.08$~mag and no
host reddening \citep[$-19.24$ and $-18.94$,
respectively;][]{Li01:00cx}.

\subsubsection{The $B$ and $V$ Bands}

While we do not have data on the rising portion of the $B$-band light
curve of SN~2013bh, we note that SN~2000cx was found to have a similar
rise time to that of the normal Type~Ia SN~1994D
\citep{Li01:00cx}. Based on two data points, SN~2013bh has a slightly
slower decline than SN~2000cx in the $B$ band (much like the $r$ and
$i$ bands, see below), but has a very similar $V$-band light
curve. SN~2000cx was found to have a normal $B$-band decline rate
until 6~d after $B$-band maximum brightness, after which time the
decline slowed \citep{Li01:00cx}. 

\subsubsection{The $r$ Band}

\begin{figure*}
\centering$
\begin{array}{c}
\includegraphics[width=4in]{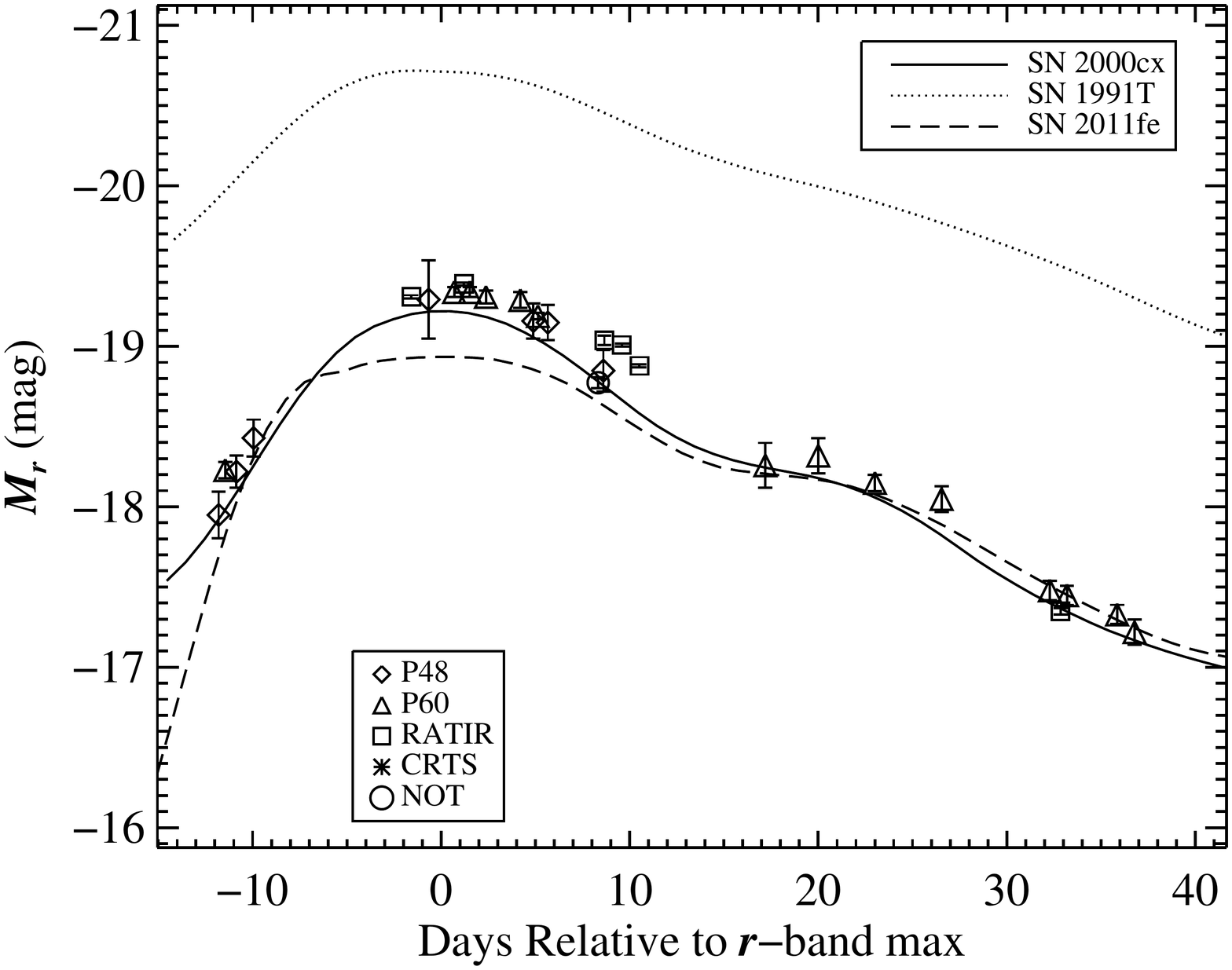} \\
\includegraphics[width=4in]{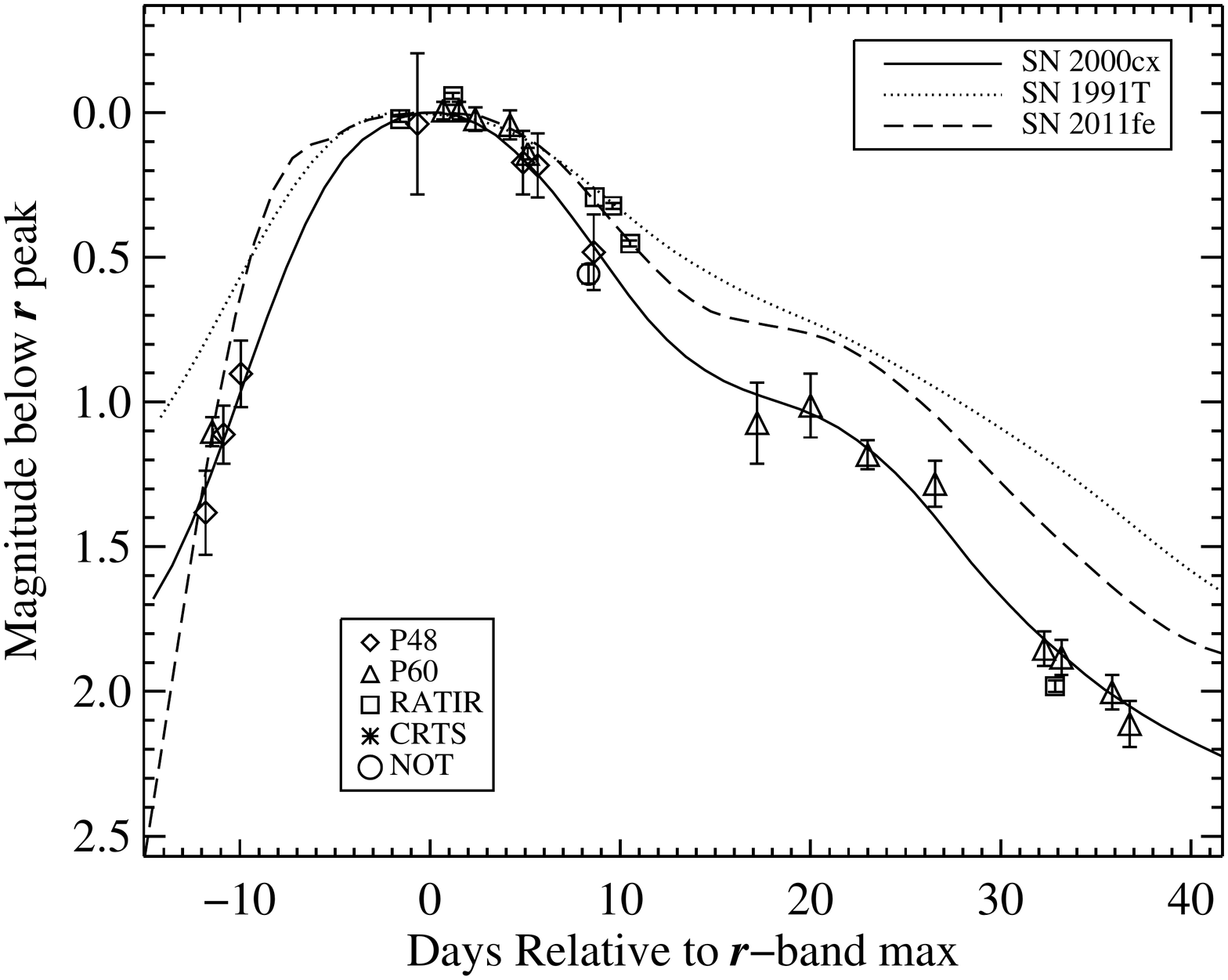}
\end{array}$
\caption{The $r$-band light curve of SN~2013bh with some comparison
  objects: SN~2000cx \citep[solid;][]{Li01:00cx}, the overluminous
  type~Ia SN~1991T \citep[dotted;][]{Lira98}, and the normal type~Ia
  SN~2011fe \citep[dashed;][]{Vinko12}. All $R$-band photometry
  has been converted to $r$-band magnitudes using conversions found in 
  \citet{Jordi06}. All data have been deredshifted and dereddened by values
  given in their respective references. Absolute $r$-band magnitudes
  are shown ({\it top}) as well a shifted version of the data such
  that all objects have the same peak magnitude ({\it
    bottom}).}\label{f:lc_comp}
\end{figure*}

The best observed bandpass of SN~2013bh is the $r$ band, and in
Figure~\ref{f:lc_comp} we present its $r$-band light curve, along with
data from various comparison objects: SN~2000cx
\citep[solid;][]{Li01:00cx}, the overluminous type~Ia SN~1991T
\citep[dotted;][]{Lira98}, and the normal type~Ia SN~2011fe
\citep[dashed;][]{Vinko12}. All comparison objects have had their
published $R$-band data converted to $r$-band magnitudes using the
conversions in \citet{Jordi06} and have been deredshifted and
dereddened by values given in their respective references. The top
panel of Figure~\ref{f:lc_comp} shows absolute $r$-band magnitudes
while the bottom panel is the same data, but shifted such that all
objects have the same peak magnitude.

Once again, the light curve of SN~2013bh matches well to
that of SN~2000cx and both are slightly more luminous at peak (in the
$r$-band) than SN~2011fe, but less luminous than SN~1991T. The
$r$-band is the only band for which we have pre-maximum data of
SN~2013bh. Based on these observations, it appears that SN~2013bh has
a relatively normal rise time, similar to that of SN~2000cx.

The $r$ band of SN~2013bh shows a plateau that begins
between 12 and 17~d after $r$-band maximum brightness, similar to
SN~2000cx for which the plateau begins at \about15~d after maximum,
which is later than more normal SNe~Ia \citep[\about12~d for SN~1994D
and \about14~d for SN~2011fe,][respectively]{Li01:00cx,Vinko12}. Since
the plateau begins later and the secondary $r$-band maximum is weaker
in SNe~2013bh and 2000cx that in other SNe~Ia, the amount these
objects have faded shortly after maximum brightness is larger than
that of the comparison objects shown in Figure~\ref{f:lc_comp}. As
mentioned above, SN~2013bh has a slightly slower $r$-band decline rate
after maximum brightness than SN~2000cx, but it is clear that both are
faster than the other objects in the Figure. Finally, the $r$-band
plateau in SN~2013bh ends between 23 and 27~d past maximum, again
similar to SN~2000cx.

\subsubsection{The $iZY\!JH$ Bands}

The $i$-band data of SN~2000cx was normal until about 7~d after
$r$-band maximum brightness at which time it began to decline faster
than normal SNe~Ia. At 14~d past maximum it had faded \about1.1~mag
below peak after which it re-brightened somewhat \citep{Li01:00cx}. It
is likely that SN~2013bh also behaved this way, but unfortunately, we
do not have $i$-band data at this epoch. We do see evidence, however, 
that SN~2013bh (like SN~2000cx) has a weak secondary maximum, as
compared to more normal SNe~Ia, in the $i$ band (just like the $r$
band). These secondary maxima are usually associated with \ion{Fe}{III}
recombining to form \ion{Fe}{II} as the SN ejecta cools
\citep{Pinto00}. Weak secondary maxima indicate that the SN
photosphere remains relatively hot through these epochs, which is
consistent with the blackbody temperatures and strong \ion{Fe}{III}
absorptions seen in SN~2013bh at these epochs (see
Section~\ref{ss:synapps}).

Moving further into the IR, the $Z$ and $Y$ data of SN~2013bh
also show weak secondary maxima as compared to other SNe~Ia, which is
similar to the weak secondary maxima in $JHK$ data of SN~2000cx
\citep{Candia03}. SN~2000cx, and many other SNe~Ia, have been found to
have $M_H \approx -17.9$~mag at 10~d after $B$-band maximum
\citep{Candia03}, which is consistent with our observed upper limit of
$M_H > -18.9$~mag at this epoch. Furthermore, \citet{Candia03}
observed a deep dip in the $J$-band light curve of SN~2000cx at about
13~d past maximum brightness and this is certainly consistent with our
upper limits for SN~2013bh.

\subsection{Colour Curves}\label{ss:colour}

\begin{figure*}
\centering$
\begin{array}{cc}
\includegraphics[width=3.2in]{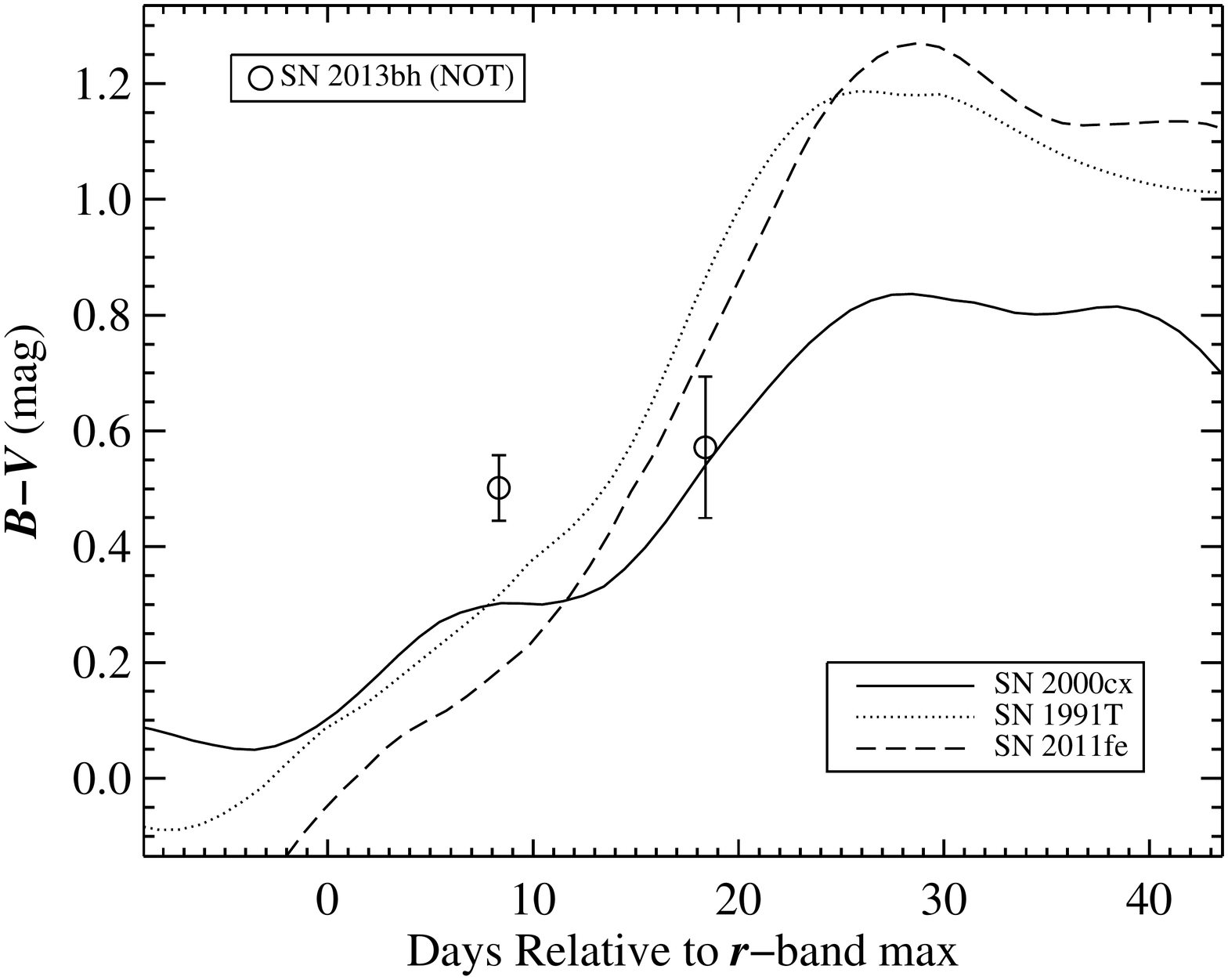} &
\includegraphics[width=3.2in]{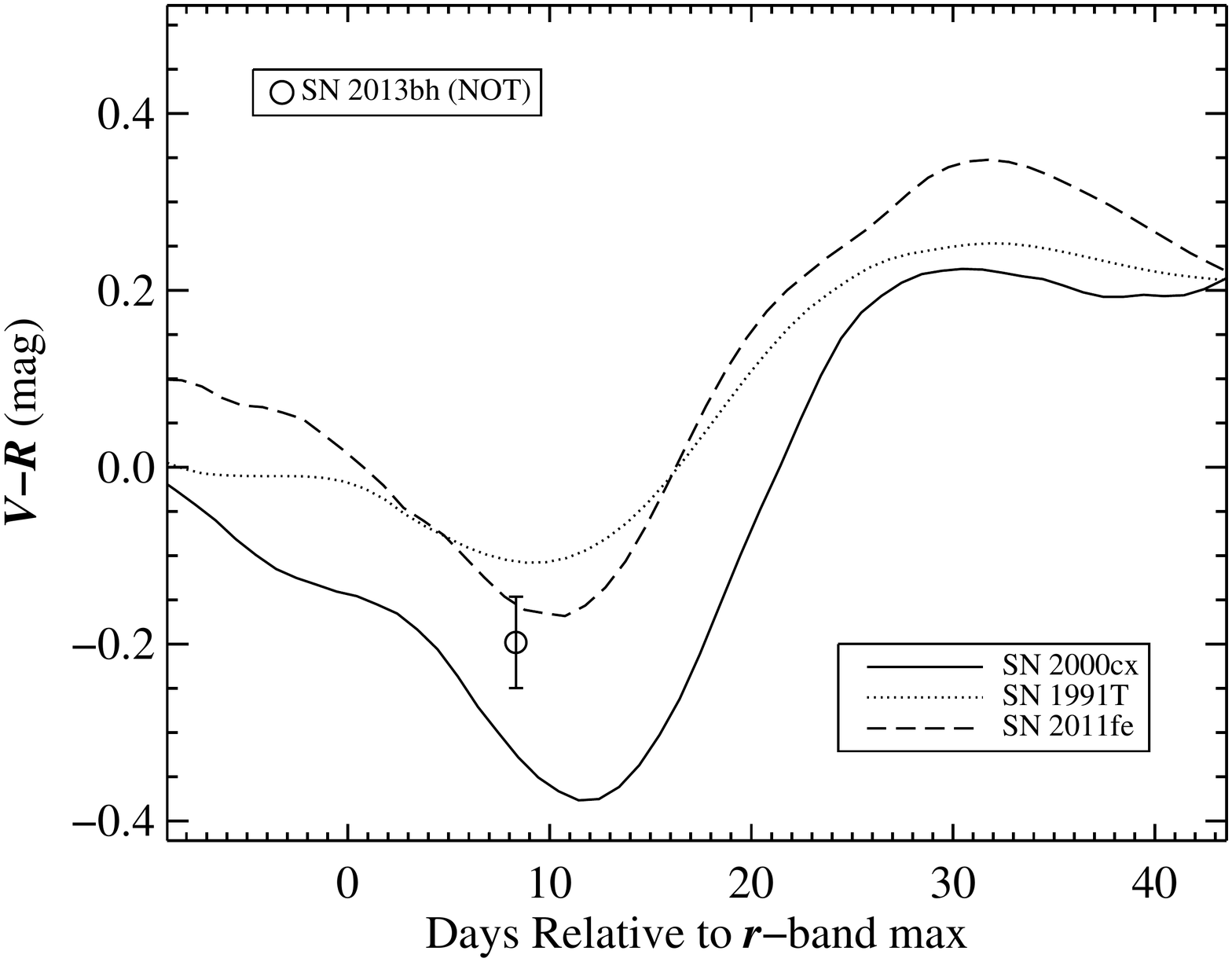} \\
\includegraphics[width=3.2in]{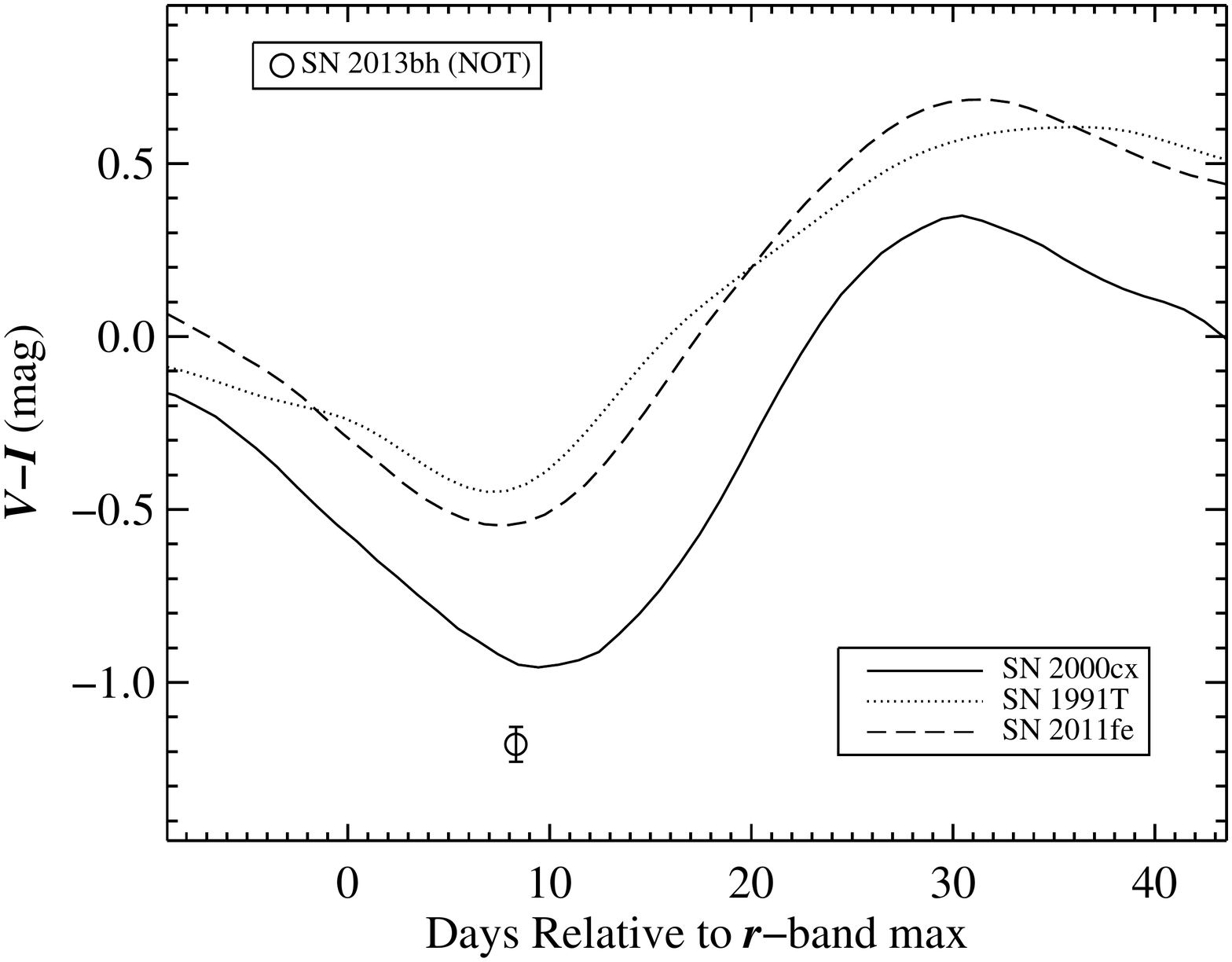} &
\includegraphics[width=3.2in]{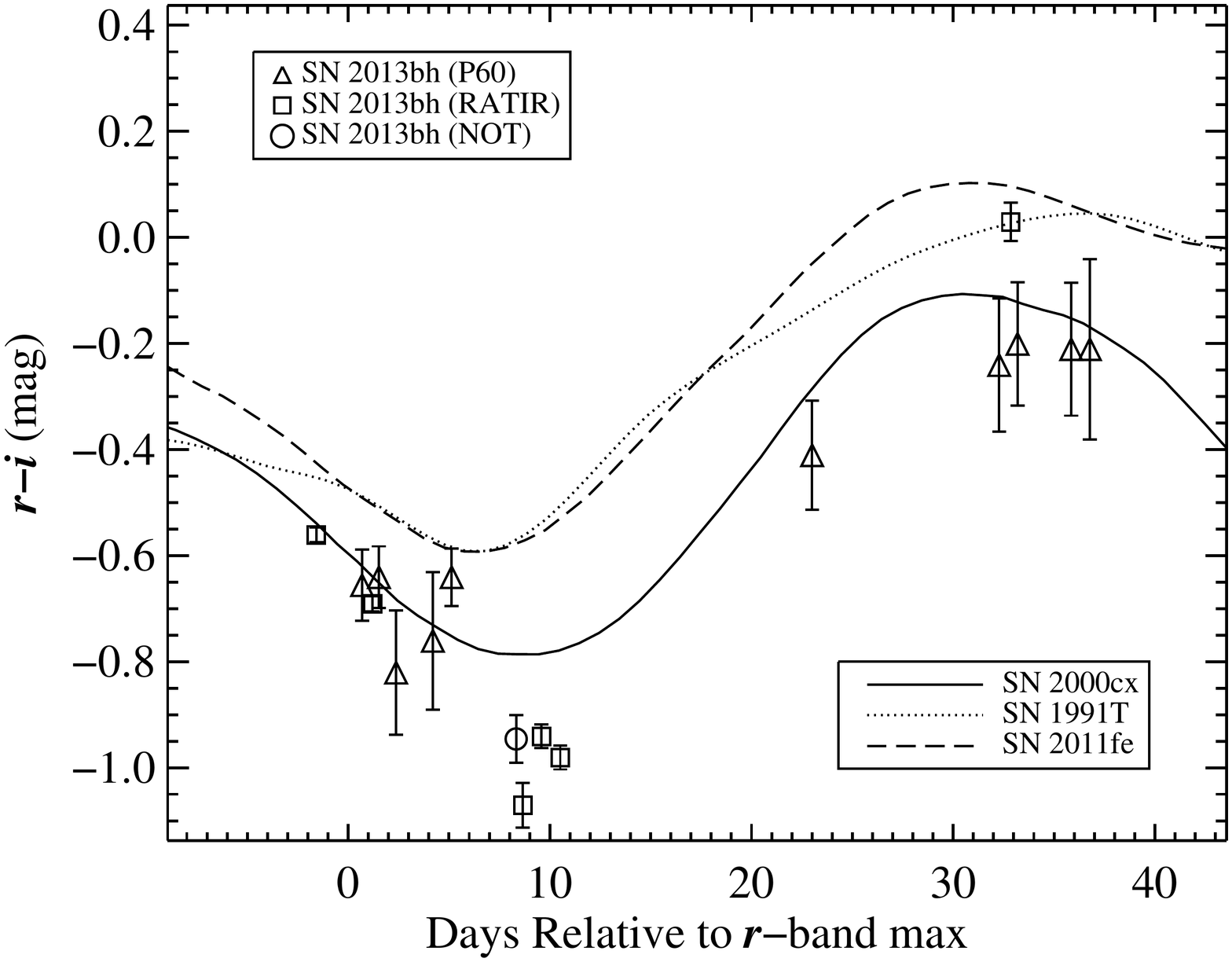} \\
\end{array}$
\caption{Colour curves of SN~2013bh and the same comparison objects
  used in Figure~\ref{f:lc_comp}. All data have been
  deredshifted and corrected for extinction using the reddening values
  provided in their respective references.}\label{f:colour}
\end{figure*}

In Figure~\ref{f:colour} we plot the colour evolution of SN~2013bh,
along with the same comparison objects that were shown in
Figure~\ref{f:lc_comp}. As above, all objects have been corrected for
Galactic extinction using the dust maps of \cite{Schlegel98} and
information in their respective references.

The $B-V$ colour of SN~2013bh is redder than all of the
comparison objects at \about8~d after $r$-band maximum brightness, but
then becomes nearly as blue as SN~2000cx by \about18~d after
maximum. The $B-V$ colour of SN~2000cx was also seen to be quite red at
early times, becoming extremely blue by 2 weeks after maximum. This
was due to a plateau in the colour curve at $B-V = 0.3$~mag for $6 < t
< 15$~d \citep{Li01:00cx,Candia03}. Even though SN~2013bh data are 
sparse, the observations are consistent with a plateau at $B-V
\approx 0.4--0.5$~mag at similar epochs as the one seen in SN~2000cx,
though perhaps a bit later.

The single $V-R$ observation of SN~2013bh from \about8~d after maximum
brightness is slightly redder than the $V-R$ colour of SN~2000cx
at the same epoch. SN~2000cx (and presumably SN~2013bh as well) is
bluer than normal and overluminous SNe~Ia at all epochs studied herein
and shows a deep blue dip at \about12~d after maximum \citep[as seen
in the upper-right panel of Figure~\ref{f:colour} and
][]{Li01:00cx}. In $V-I$ it was also seen that SN~2000cx was bluer
than other SNe~Ia and again showed a deep blue dip, though in this
colour the dip is a few days earlier than in $V-R$. We see a similar
behavior in our one $V-I$ data point for SN~2013bh, though it is
significantly (\about0.3~mag) bluer than even SN~2000cx.

The $r-i$ colours of SN~2013bh, for which we have a handful of data
points, are found to match SN~2000cx near $r$-band maximum
brightness. Soon after maximum, however, SN~2013bh evolves to be bluer
than SN~2000cx in $r-i$. The minimum in the $r-i$ colour curve of
SN~2013bh occurs at \about9~d after maximum and is about 0.2~mag bluer
than that of SN~2000cx. At later epochs, SN~2013bh remains
\about0.1~mag bluer than SN~2000cx. Note that the majority of the
difference in $r-i$ of SN~2013bh at about 1 month past maximum as
measured from P60 and RATIR data comes from the $i$-band photometry,
though it is still consistent at about the 2$\sigma$
level. Furthermore, no instrumental response corrections have been
applied to our data and, as mentioned above, no host-galaxy
subtraction was performed.

In summary, based on limited data, the colours of SN~2013bh tend to
follow those of SN~2000cx, and certainly more closely than either
SN~1991T or SN~2011fe. However, SN~2013bh appears to have more extreme
colours than SN~2000cx. In general SN~2013bh seems to be redder than
SN~2000cx in the bluest colours, but bluer than SN~2000cx in the
reddest bands. This implies an optical continuum that peaks redder
than SN~2000cx, thus SN~2013bh seems to have a lower blackbody
temperature which is consistent with our spectral models (see
Section~\ref{ss:synapps}). 

\subsection{Spectra}\label{ss:spectra}

\begin{figure*}
\centering$
\begin{array}{c}
\includegraphics[width=5.8in]{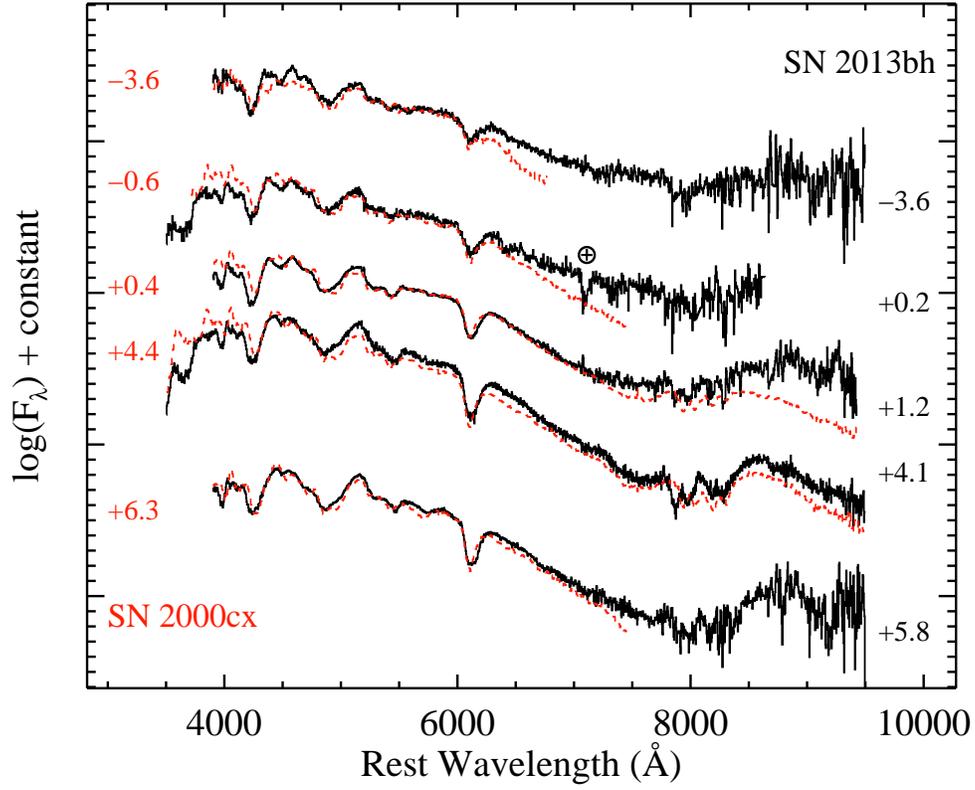} \\
\includegraphics[width=5.8in]{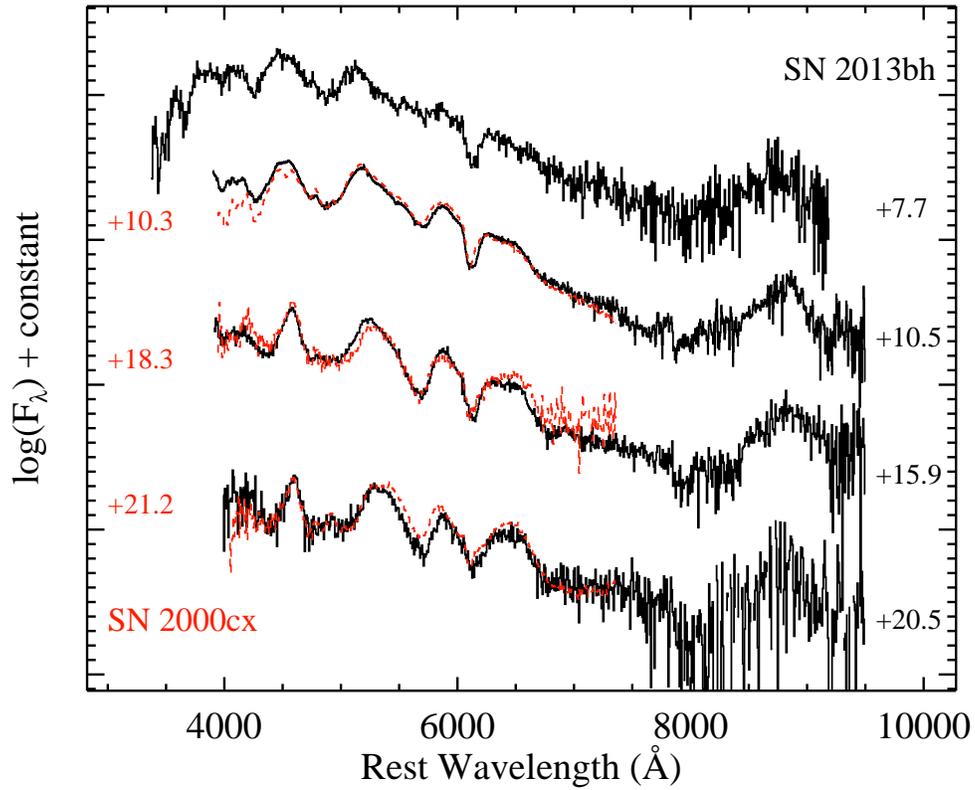}
\end{array}$
\caption{Spectra of SN~2013bh (solid black) and SN~2000cx \citep[dashed
  red, taken from][]{Li01:00cx,Matheson08}, both labelled with their
  age relative to $r$-band maximum brightness. Telluric absorption in
  the maximum light spectrum is marked by the Earth symbol ($\earth$)
  and both objects have been deredshifted and
  dereddened.}\label{f:spec}
\end{figure*}

All of our spectra of SN~2013bh are shown in black in
Figure~\ref{f:spec} and are labelled with their age relative to
$r$-band maximum brightness. The over-plotted red, dashed curves are 
spectra of SN~2000cx at similar epochs to our observations of
SN~2013bh, also labelled by their age relative to $r$-band maximum
brightness \citep[converted from the date of $R$-band maximum
brightness as determined by][]{Li01:00cx}. The SN~2000cx comparison
data come from the in-depth study of that object by \citet{Li01:00cx}
and the large CfA SN~Ia spectral dataset presented by
\citet{Matheson08}. Both objects have been deredshifted and
dereddened. The spectral match between SN~2013bh and SN~2000cx at all
epochs studied in this work is remarkable. We do note, however, that
SN~2000cx seems to have a slightly bluer continuum than SN~2013bh,
which is indicative of a higher blackbody temperature, once again
consistent with our spectral models (see Section~\ref{ss:synapps}).  

We assumed above no reddening from the host galaxy of SN~2013bh. This
is due mainly to the facts that we see no evidence of absorption from
\ion{Na}{I}~D and because it was found in the outskirts of its host
(much like SN~2000cx). In our highest S/N 
spectrum from 4~d after $r$-band maximum brightness, we find a
2$\sigma$ upper limit of 0.2~\AA\ for the equivalent width (EW) of
\ion{Na}{I}~D absorption. Using the empirical relation of
\citet{Poznanski11}, this converts to $E(B-V)_{\textrm host} <
0.006$~mag, which is negligible (especially given the uncertainties in
the conversion, i.e. \about0.3~mag). Also supporting no host reddening
is the location of SN~2013bh in its host galaxy. It went off
\about24~kpc (projected linear distance) from the centre of
SDSS~J150214.17+103843.6, which is equivalent to \about3 Petrosian
radii \citep{Petrosian76}. See Section~\ref{ss:host} for more
information on the host galaxy of SN~2013bh.

\begin{figure*}
\centering
\includegraphics[width=5.5in]{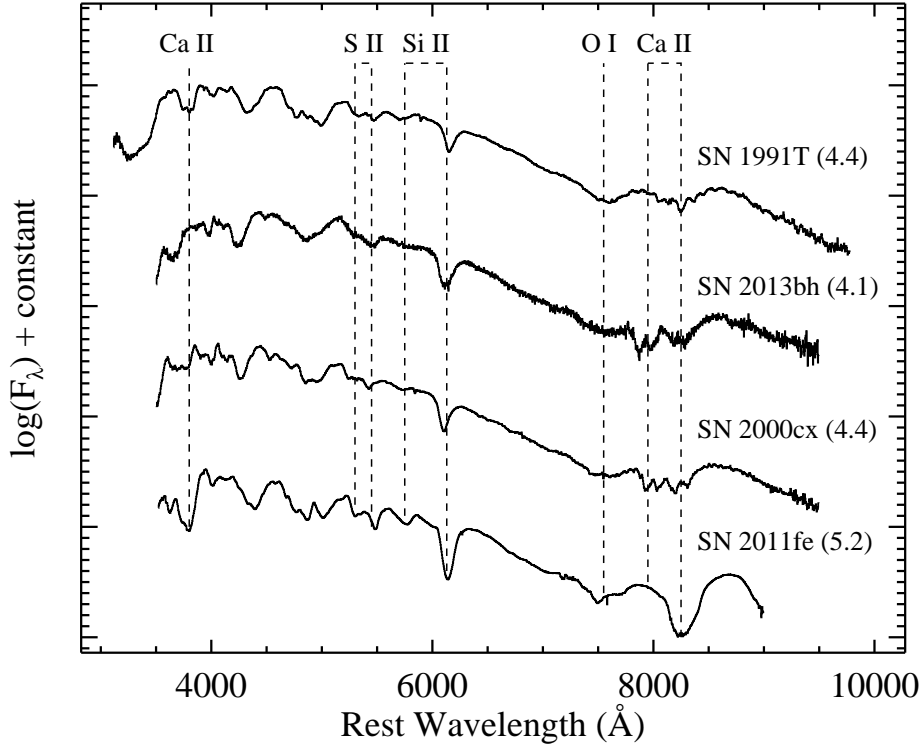}
\caption{A spectrum of SN~2013bh along with comparisons to other SNe~Ia:
  SN~1991T \citep{Filippenko92:91T}, SN~2000cx \citep{Li01:00cx}, and
  SN~2011fe \citep{Parrent12}. Each spectrum is labelled with its age
  relative to $r/R$-band maximum brightness and the data have all been
  deredshifted and dereddened. Significant spectral features have been 
  identified.}\label{f:spec_comp}
\end{figure*}

In Figure~\ref{f:spec_comp} we compare our highest S/N spectrum of
SN~2013bh (from \about4~d past maximum brightness) with various other
SNe~Ia at similar epochs. Shown in the Figure are its near twin,
SN~2000cx \citep{Li01:00cx}, as well as the overluminous SN~1991T
\citep{Filippenko92:91T} and the extremely normal SN~2011fe
\citep{Parrent12}. Important spectral features are labelled. SN~2011fe
stands out from the other three objects with overall much stronger
absorption features and a much broader and more blended \ion{Ca}{II}
near-IR triplet. Further differences are seen in the blue end of the
spectrum where SN~2011fe is dominated by absorption from blends of
\ion{Fe}{II} and \ion{Mg}{II} lines, while the complex profiles in the
other objects are due in large part to \ion{Fe}{III} and \ion{Ti}{II}
(see \S\ref{ss:synapps} for further information).

Even though SNe~2013bh and 2000cx bear some resemblance to SN~1991T,
there are significant differences as well. The \ion{O}{I} line is more
prominent in SN~1991T than in either SNe~2013bh and 2000cx, where it
is almost non-existent. This (along with a lack of C absorption) may
be an indication of more complete WD burning in SNe~2013bh and 2000cx,
which is consistent with the relatively large amount of $^{56}$Ni
inferred for these objects (see \S~\ref{ss:host}). SNe~2013bh and
2000cx both also show significant absorption from HVFs of \ion{Ti}{II}
while SN~1991T lacks these features in the blue end of the
spectrum. In addition, just blueward of the \ion{Ca}{II} near-IR
triplet in these latter two objects, a second absorption component
from HVFs of \ion{Ca}{II} is easily seen and is conspicuously absent
from SN~1991T.

\subsection{Spectral Fits}\label{ss:synapps}

The spectrum-synthesis code {\tt SYNAPPS} \citep{Thomas11:synapps} was
employed to help identify the species present in our spectra of
SN~2013bh. {\tt SYNAPPS} (and its modeling kernel {\tt SYN++}) is
derived from {\tt SYNOW} \citep{Synow}, which computes synthetic
spectra of SNe in the photospheric phase using the Sobolev
approximation \citep{Sobolev60,Castor70,Jeffery89}. {\tt SYNAPPS} is
capable of varying a large number of parameters automatically, thus
finding an optimum fit via $\chi^2$-minimization.

In {\tt SYNAPPS} models the spectral lines are assumed to form via
resonance scattering above a sharp photosphere. The location of the
photosphere is expressed in velocity coordinates as $v_\textrm{ph}$
(in \kms) taking into account the homologous expansion of the SN
ejecta. Consequently, for a particular ion, the minimum and maximum
velocity coordinates of the line-forming region are denoted with
$v_\textrm{min}$ and $v_\textrm{max}$ (both in \kms), respectively. If
$v_\textrm{min} \ga ~v_\textrm{ph}$, then the line forming region is
considered ``detached'' from the photosphere.

The optical depths for each transition of a given species are
controlled with two additional parameters: the optical depth of a
``reference line,'' $\tau_\textrm{ref}$ (which is usually the
strongest line in the optical band, defined internally within the
code), and the e-folding width, $v_e$ (in \kms), of the optical depth
profile above the photosphere (assumed to be an exponential function
in {\tt SYNAPPS}). For other features, the optical depth is computed
relative to the reference line assuming Boltzmann-excitation
(i.e. local thermodynamic equilibrium) using an excitation temperature
$T_\textrm{exc}$ (in K). Non-LTE effects are mimicked partly by
allowing different $T_\textrm{exc}$ values for each species, all of
which can be different from the photospheric temperature
$T_\textrm{phot}$. The latter is used only in computing the blackbody
radiation emitted by the photosphere. Since the number of species
appearing in a typical SN spectrum is usually between 5 and 10, and
each ion has 5 tunable parameters, the number of adjustable parameters
in a {\tt SYNAPPS} session can be substantial.

The spectral sequence of SN~2013bh in Figure~\ref{f:spec} suggests
that the optical spectra shortly before and after maximum brightness
are formed by the same set of ions, and the spectral evolution can be
modeled by simply tuning the optical depth and other parameters of the
same species. This is supported by the similar photospheric
temperature of the pre-maximum and post-maximum spectra, which
suggests that the excitation of ions should not change drastically
near maximum.

\begin{figure*}
\centering
\includegraphics[width=5.5in]{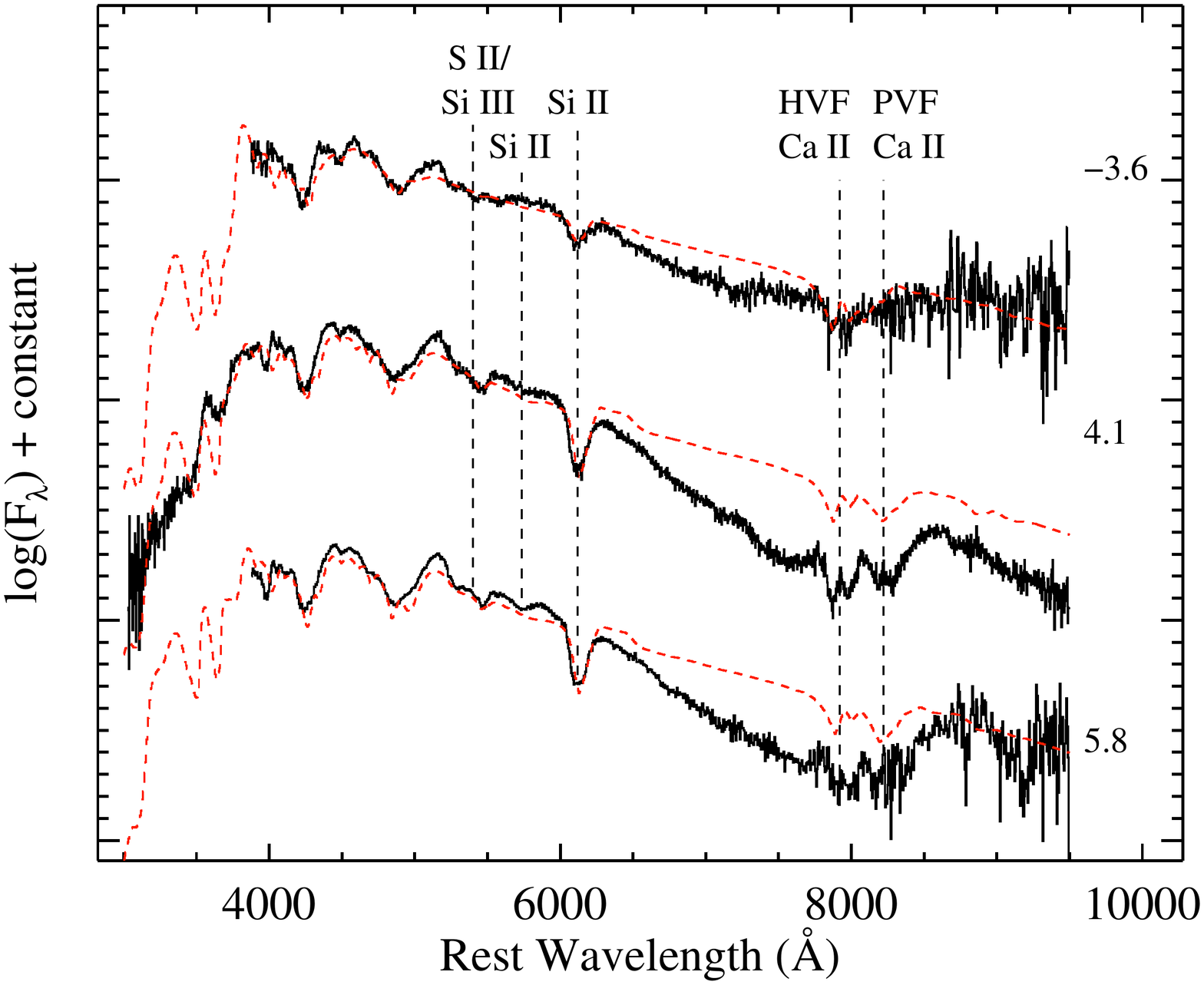}
\caption{Observed spectra of SN~2013bh (solid black) and {\tt SYNAPPS} 
  fits to the data (dashed red). Each spectrum is labelled with
  its age relative to $r$-band maximum brightness and the data have
  all been deredshifted and dereddened. Major spectral features are
  identified.}\label{f:synapps}
\end{figure*}

A model containing the usual composition of a SN~Ia was constructed to
fit the spectra of SN~2013bh and is based on the model for SN~2000cx
by \citet{Branch04:00cx} and \citet{Thomas04}. 
The model was initially fit to the 4.1~d spectrum, then it
was optimized for the $-3.6$ and +6.3~d spectra. {\tt SYNAPPS} fits to
these three SN~2013bh spectra are shown in Figure~\ref{f:synapps}. The
data are shown as solid black, while the models are shown as dashed
red. Major spectral features have been labelled. The parameters of our
{\tt SYNAPPS} fit to the 4.1~d spectrum are listed in
Table~\ref{t:modelI}. 

\begin{table*}
\begin{center}
\caption{{\tt SYNAPPS} Model Parameters for the 4.1~d Spectrum\label{t:modelI}}
\begin{tabular}{lccccc}
\hline \hline
Ion & $\log \tau_\textrm{ref}$ & $v_\textrm{min}$ & $v_\textrm{max}$ & $v_e$ & $T_\textrm{exc}$ \\
      &                   & ($10^3$~\kms) &  ($10^3$~\kms) &  ($10^3$~\kms) & ($10^3$ K) \\
\hline
\ion{Si}{II} & $\phantom{-}$1.57 & 12.0 & 40.0 & 1.0 & $\phantom{1}$8.8 \\
\ion{Si}{III} & $\phantom{-}$1.34 & 16.0 & 40.0 & 2.0 & 13.0 \\
\ion{S}{II} & $-2.00$ & 12.0 & 40.0 & 1.0 & $\phantom{1}$5.2 \\
\ion{Ca}{II} (PVF) & $\phantom{-}$0.80 & 12.0 & 40.0 & 4.8 & 12.0 \\
\ion{Ca}{II} (HVF) & $\phantom{-}$2.90 & 25.0 & 40.0 & 3.0 & 12.0 \\
\ion{Ti}{II} & $\phantom{-}$2.09 & 21.0 & 40.0 & 2.0 & $\phantom{1}$5.5 \\
\ion{Fe}{II} & $\phantom{-}$0.94 & 21.2 & 40.0 & 4.0 & $\phantom{1}$5.0 \\
\ion{Fe}{III} & $\phantom{-}$0.03 & 11.0 & 40.0 & 2.0 & 12.0 \\ 
\ion{Co}{II} & $\phantom{-}$0.60 & 11.0 & 40.0 & 2.0 & $\phantom{1}$6.3 \\
\ion{Ni}{II} & $\phantom{-}$1.00 & 11.0 & 40.0 & 2.0 & $\phantom{1}$5.1 \\
\hline\hline
\multicolumn{6}{l}{$v_\textrm{ph} = 11000$~\kms, $T_\textrm{phot} = 11000$~K.} \\
\hline\hline
\end{tabular}
\end{center}
\end{table*}


The model includes PVFs of \ion{Ca}{II}, \ion{Si}{II}, \ion{S}{II},
\ion{Fe}{III}, \ion{Co}{II}, and \ion{Ni}{II}, in addition to
detached, HVFs of \ion{Ca}{II}, \ion{Si}{III}, \ion{Ti}{II}, and
\ion{Fe}{II}. The photospheric velocity is $v_\textrm{ph} \approx
11000$~\kms for all modeled spectra and the detached, HVFs span a
range between 20000 and 24000~\kms. These modeled velocities are fully
consistent with what is calculated from direct measurements of the
spectral features themselves (\S~\ref{ss:vels}), including a
\ion{Si}{II} velocity plateau seen in our SN~Ia-like {\tt SYNAPPS}
fits. From Figure~\ref{f:synapps}, it is clear that most of the
spectral features between 3000 and 6300~\AA\ can be fit successfully
with the models. As also seen in other, more normal SNe~Ia at similar
epochs \citep[e.g.,][]{Silverman12:BSNIPI}, redward of \about6300~\AA,
the continuum is significantly depressed relative to the blackbody 
continuum, which cannot be modeled self-consistently within the
framework of {\tt SYNAPPS}. Therefore, this region has been omitted
from the automatic fitting process.

We identify \ion{Ca}{II} via the near-IR triplet (at \about8000~\AA)
as well as \ion{Ca}{II}~H\&K (at \about3650~\AA). The strong,
unblended feature near 6150~\AA\ is due to \ion{Si}{II} $\lambda$6355
and the feature near 5480~\AA\ is from \ion{S}{II}. The model is able
to fit the observed feature near 4000~\AA, which was identified in
SN~2000cx as a blended HVF from \ion{Ti}{II} by \citet{Branch04:00cx}
and could also be due to \ion{Si}{II} $\lambda$4130, despite the
presence of these ions in our model. Note that the feature at
~4500~\AA, suspected to be a HVF of H$\beta$ in SN~2000cx by
\citet{Branch04:00cx}, can be fit adequately as a blend of iron-group
elements. 


As mentioned above, our {\tt SYNAPPS} model of SN~2013bh is extremely
similar to models of SN~2000cx. \citet{Li01:00cx} found 
that \ion{Si}{II} $\lambda$6355 strengthened with time in SN~2000cx
while \ion{Fe}{III} remained strong through maximum brightness, which is
identical to what we find for SN~2013bh. \ion{S}{II} was found to be
weak in both objects, but vanished quicker in SN~2013bh than
SN~2000cx. HVFs of \ion{Fe}{II} occurred in both objects, but we find
evidence of only PVFs for \ion{Fe}{III} and \ion{Si}{II} while SN~2000cx
showed evidence for HVFs of these ions
\citep{Li01:00cx,Thomas04}. Evidence for \ion{O}{I} and \ion{Mg}{II}
absorption in SN~2000cx was presented by \citet{Thomas04} and
\citet{Branch04:00cx}, respectively, though our {\tt SYNAPPS} fits do
not support the same claim for SN~2013bh. Models in both of
those studies do indicate \ion{Ca}{II} velocities in SN~2000cx that match
what we find for SN~2013bh quite well. In addition,
\citet{Branch04:00cx} note significant absorption due to HVFs of
\ion{Ti}{II} in their models of SN~2000cx and we find similar results
for our models of SN~2013bh.

For epochs studied herein, the photospheric temperatures inferred from
our {\tt SYNAPPS} fits to SN~2013bh are \about11000~K. This is higher
than more normal SNe~Ia, though slightly cooler than what was found
for SN~2000cx \citep[12000~K;][]{Thomas04}. Note, however, that at
slightly earlier epochs (about 1 week before maximum brightness), a
much higher blackbody temperature for SN~2000cx was inferred from
near-IR spectra \citep[20000-25000~K;][]{Rudy02}.

\subsection{Expansion Velocities}\label{ss:vels}

For a few of the spectral features in SN~2013bh (and SN~2000cx) which
are easily-identifiable and relatively well-separated, the expansion
velocities are measured. Here we investigate the \ion{Si}{II}
$\lambda$6355 feature, the \ion{S}{II} ``W'' feature,\footnote{The two
  broad absorptions that make up the \ion{S}{II}  ``W'' are fit using
  a single spline, but we calculate the expansion velocity of the
  absorption complex using the minimum of the bluer of the two
  features relative to its rest wavelength.} the \ion{Ca}{II}~H\&K
feature, and the \ion{Ca}{II} near-IR triplet. For the \ion{Si}{II} and
\ion{S}{II} features, the method of velocity determination is the same
as that used in \citet{Silverman12:BSNIPII}, but, briefly, consists of:
defining two endpoints on either side of the feature of interest,
fitting a linear pseudo-continuum between those endpoints, fitting a
spline function to the data between the endpoints, and using the
minimum of the spline fit to calculate an expansion velocity for the
feature. 

For the \ion{Ca}{II} features, however, we employed a slightly
different fitting method. Similar to the method outlined above, we
begin by choosing endpoints for the spectral feature in question and,
using these endpoints, define and then subtract from the data a linear
pseudo-continuum. Then a non-linear least squares fitter is used to
simultaneously fit multiple Gaussian components to each component of
the feature. The \ion{Ca}{II}~H\&K feature (\ion{Ca}{II} near-IR
triplet) consists of 2 (3) distinct components. Given that the
relative fluxes (set by the $gf$ weights) and rest wavelengths of each
component are known and since we require that all components in a
given fit have the same Gaussian width, each \ion{Ca}{II} fit contains
only 3 free parameters: the Gaussian height, width, and centroid of
the strongest component. By-eye initial estimates of these parameters
for each fit are used as inputs to the fitting routine.

For the \ion{Ca}{II}~H\&K profile, we also allow for the existence of
absorption due to \ion{Si}{II} $\lambda$3858 at a similar velocity to
\ion{Si}{II} $\lambda$6355 \citep[e.g.,][]{Foley13:cahk}. We find that
\ion{Si}{II} $\lambda$3858 is extremely weak (or non-existent) in our
spectra of SN~2013bh that cover this wavelength range. We do, however,
find evidence for \ion{Si}{II} $\lambda$3858 absorption in SN~2000cx
at velocities consistent with \ion{Si}{II} $\lambda$6355.

\begin{figure*}
\centering$
\begin{array}{c}
\includegraphics[width=4.6in]{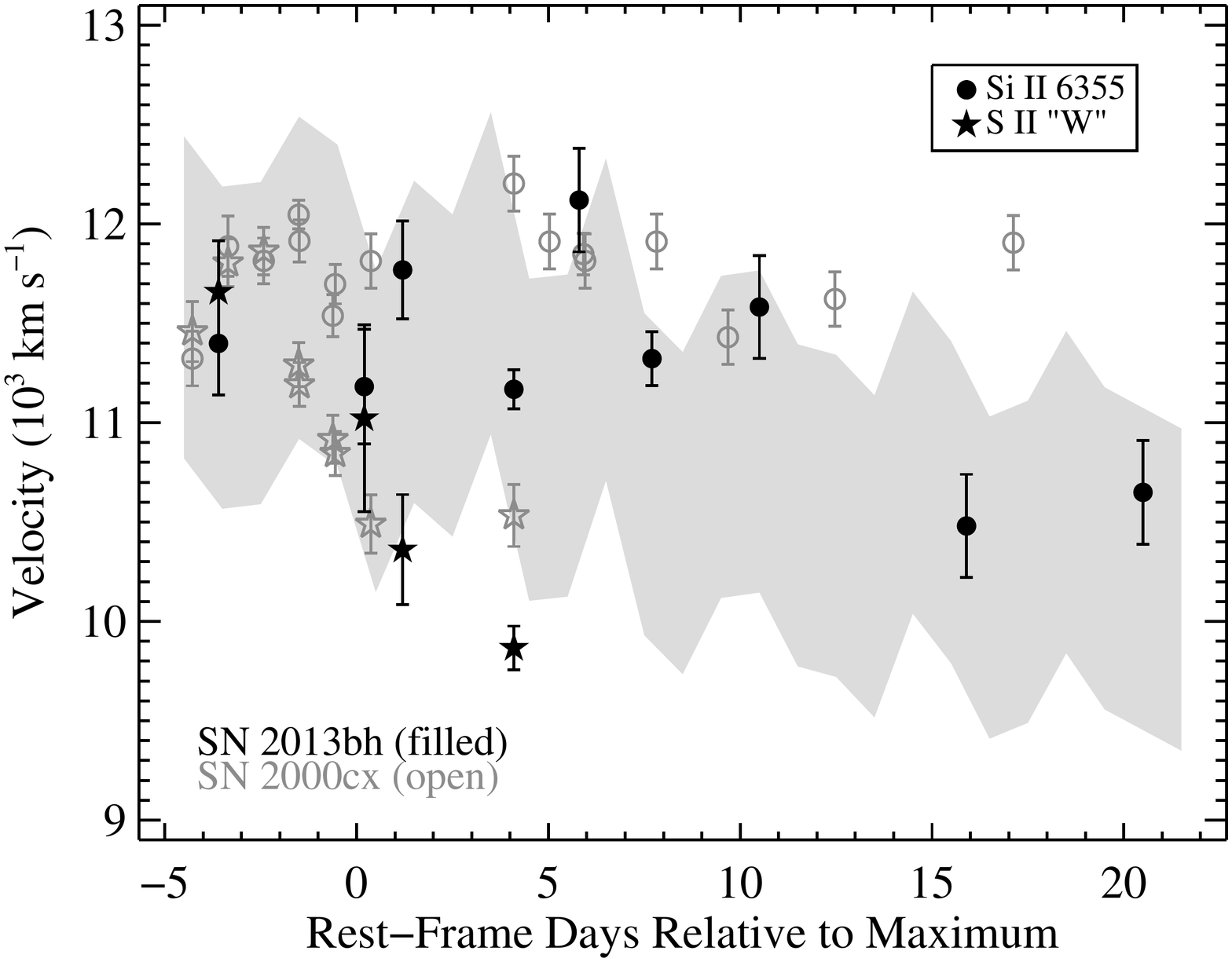} \\
\includegraphics[width=4.6in]{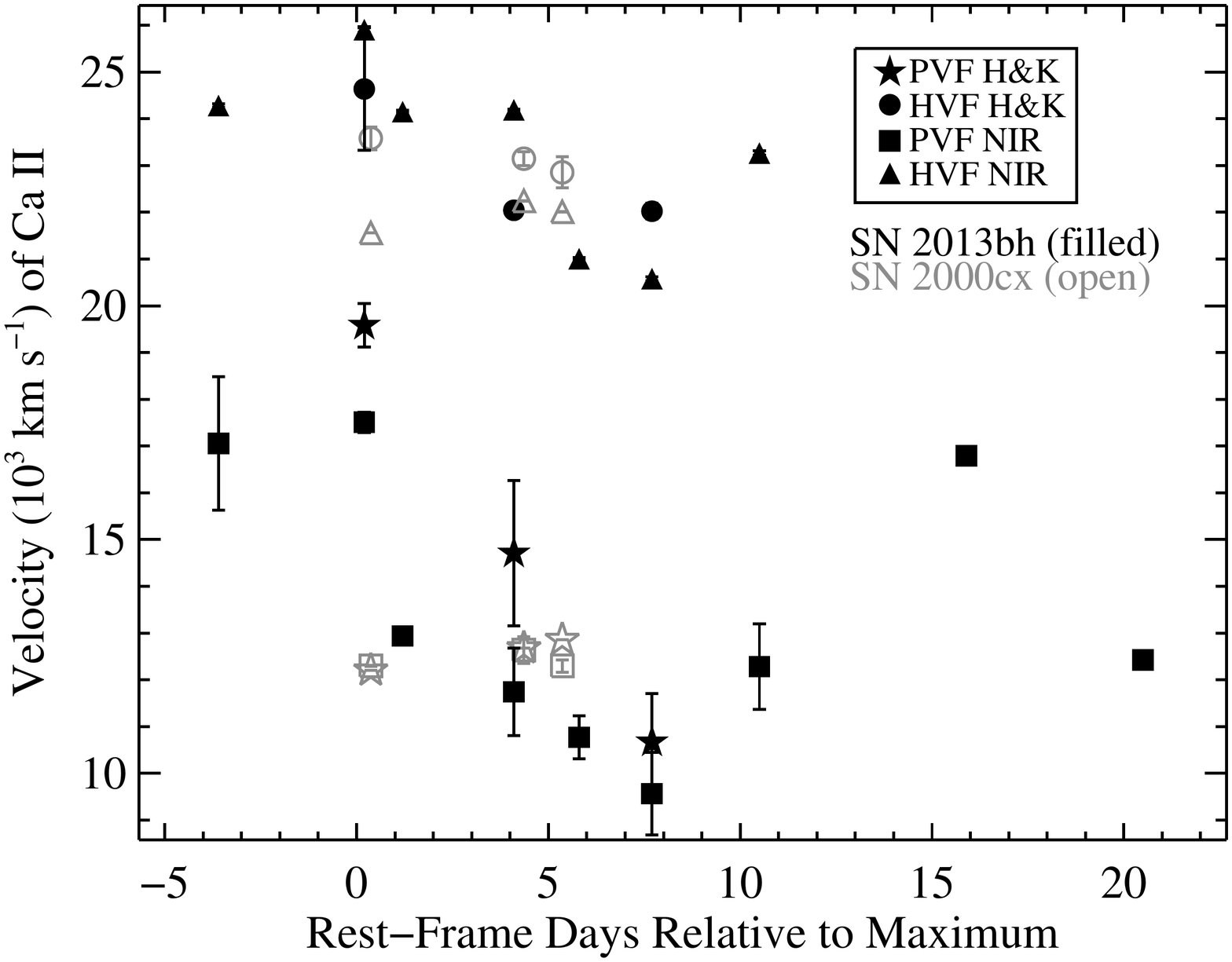}
\end{array}$
\caption{The temporal evolution of the expansion velocities (relative
  to $r$-band maximum brightness) of SN~2013bh ({\it filled black})
  and SN~2000cx ({\it open grey}). The top panel shows velocities of
  the \ion{Si}{II} $\lambda$6355 ({\it circles}) and \ion{S}{II}
  ``W'' features ({\it stars}). The bottom panel
  shows velocities of PVFs and HVFs of \ion{Ca}{II}~H\&K ({\it stars}
  and {\it circles}, respectively) and PVFs and HVFs of the
  \ion{Ca}{II} near-IR triplet ({\it squares} and {\it triangles},
  respectively). The light grey swath in the top panel is the
  1$\sigma$ region around the average \ion{Si}{II} $\lambda$6355
  velocity of normal SNe~Ia as determined by 
  the entire BSNIP dataset. Note that the HVFs and PVFs of the
  \ion{Ca}{II} near-IR triplet in SN~2013bh become blended by 16~d
  after maximum brightness and thus are shown as a single (black
  filled square) data point from then on.}\label{f:lines}
\end{figure*}

All of the measured velocities for SN~2013bh are consistent with our
{\tt SYNAPPS} fits and are plotted in Figure~\ref{f:lines} ({\it
  filled black}), relative to $r$-band maximum brightness. We  also
plot ({\it open grey}) our re-measured velocities of SN~2000cx (also
relative to $r$-band maximum) using spectra presented in
\citet{Li01:00cx} and \citet{Matheson08}. Note that the velocities for
SN~2000cx calculated herein are consistent with previous work
\citep{Li01:00cx,Thomas04,Branch04:00cx}.

The upper panel of Figure~\ref{f:lines} shows the velocities of the
\ion{Si}{II} $\lambda$6355 and \ion{S}{II} ``W'' features for SNe~2013bh
and 2000cx, along with the 1$\sigma$ region ({\it light grey})
around the average \ion{Si}{II} $\lambda$6355 velocity of normal SNe~Ia
as determined by the entire Berkeley SN~Ia Program sample
\citep[BSNIP;][]{Silverman12:BSNIPII}. The
\ion{Si}{II} $\lambda$6355 velocity of SN~2013bh is initially similar
to other, more normal SNe~Ia, but shows a plateau until \about10~d
after maximum brightness, thus becoming larger than the typical SN~Ia
\ion{Si}{II} $\lambda$6355 velocity. After this epoch, the velocity
decreases some, eventually matching the more normal SNe~Ia. SN~2000cx
also shows a plateau at nearly the same velocity
(\about11500-12000~\kms) though it is much longer lived, lasting
through all epochs for which we measure a \ion{Si}{II} $\lambda$6355
velocity in this work. In fact, \citet{Li01:00cx} found that this
plateau continues to $t > 40$d. Such long-lasting \ion{Si}{II} velocity
plateaus are not unheard of, though all other examples of them are
found in extremely peculiar SNe~Ia
\citep[e.g.,][]{Scalzo10,Childress13:12fr}.

The width of the \ion{Si}{II} $\lambda$6355 feature in SN~2013bh is
larger than that of SN~2000cx, but we do not see any convincing
evidence for HVFs from \ion{Si}{II}, even though this has been seen in
other SNe~Ia 
\citep[e.g.,][]{Silverman12:12cg,Childress13:12fr,Marion13:09ig}. In
both SNe~2000cx and 2013bh, the \ion{Si}{II} $\lambda$6355 and
\ion{S}{II} ``W'' features both strengthen with time, though while the
former stays strong through all epochs considered here, the latter
begins to weaken after maximum brightness and completely disappears by
a few days after maximum.

As seen in Figure~\ref{f:spec}, the \ion{S}{II} ``W'' feature is
somewhat weaker in SN~2013bh as compared to SN~2000cx near maximum
brightness. This spectral feature is usually found to have velocities
1000-2000~\kms\ lower than the \ion{Si}{II} $\lambda$6355 feature
\citep[e.g.,][]{Silverman12:BSNIPII}, though in SNe~2013bh and 2000cx
both features have similar velocities before maximum. At this point
the velocity plummets, likely due to the feature getting progressively 
weaker and becoming blended with other ions (notably \ion{Si}{III}, see
\S~\ref{ss:synapps}).

The spectra of SN~2000cx blueward of the \ion{Ca}{II} near-IR triplet
were initially found to be quite complex and confusing
\citep{Li01:00cx}, but it was eventually realized, through spectral
modeling, that this object showed HVFs of \ion{Ca}{II} at velocities $>
20000$~\kms\ \citep{Thomas04,Branch04:00cx}. This was in addition to a
second set of \ion{Ca}{II} features at more typical velocities
(\about12000~\kms). HVFs from \ion{Ca}{II} have been seen
in many individual SNe~Ia
\citep[e.g.,][]{Silverman12:12cg,Childress13:12fr,Marion13:09ig} and
might be common in the early-time spectra of SNe~Ia
(e.g., \citealt{Mazzali05}; \citealt{Childress13}; Silverman \etal in 
preparation). Unsurprisingly, SN~2013bh once again matches 
SN~2000cx and shows strong HVFs from \ion{Ca}{II} at early times. 

The bottom panel of Figure~\ref{f:lines} shows the velocities for
SN~2013bh ({\it filled black}) and SN~2000cx ({\it open grey}) of PVFs
and HVFs of \ion{Ca}{II}~H\&K ({\it stars} and {\it circles},
respectively) and PVFs and HVFs of the \ion{Ca}{II} near-IR triplet
({\it squares} and {\it triangles}, respectively). In both objects,
through \about10~d past maximum, we measure distinct PVFs and HVFs for
both the \ion{Ca}{II}~H\&K feature and the \ion{Ca}{II} near-IR
triplet. The PVF velocities are similar to more normal SNe~Ia at these
epochs, while the HVF velocities of both objects are significantly
larger than most SNe~Ia (\citealt{Childress13}; 
Silverman \etal in preparation). By 16~d past maximum, we are only
able to measure a single \ion{Ca}{II} near-IR triplet feature in
SN~2013bh. Since the measured velocity at this epoch is nearly the
average of the HVF and PVF velocities from the previous spectrum, it
appears that the two components have become blended and thus are shown
in the Figure as a single (black filled square) data point from then
on.

The measured velocities of the PVFs for both \ion{Ca}{II} features in
SN~2000cx are nearly identical, while the HVF velocities of this
objects also match each other quite well. This is a useful sanity
check for our velocity determinations since one might expect all
\ion{Ca}{II} features in a given spectrum to have nearly equal
velocities. The velocities of the PVFs and HVFs are quite
self-consistent for SN~2013bh as well. Also note the relatively small
scatter in {\it all} PVF velocity measurements (both objects and both
spectral features), as well as the consistency of all of the HVF
velocity measurements. This once again emphasizes the similarity
between these two objects.

\begin{figure*}
\centering
\includegraphics[width=5in]{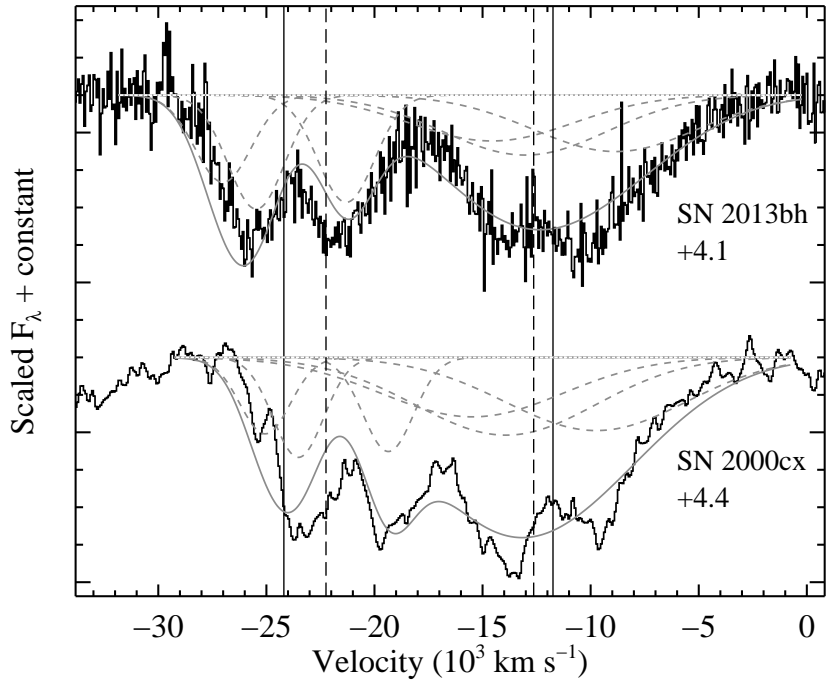}
\caption{The \ion{Ca}{II} near-IR triplet profile for SNe~2013bh ({\it
    top}) and 2000cx ({\it bottom}), after removing the linear
  pseudo-continuum (grey dotted line). The rest-frame age of each
  spectrum is listed next to the data ({\it black}). Over-plotted are
  the combined Gaussian fits ({\it solid grey}) and the individual
  Gaussian components ({\it short-dashed grey}). The HVFs are the
  left-side Gaussians while the PVFs are the right-side
  Gaussians. The \ion{Ca}{II} near-IR triplet velocity of SN~2013bh
  (SN~2000cx) is shown as the vertical solid (long-dashed)
  line.}\label{f:cair} 
\end{figure*}

To highlight the separation between the HVFs and the PVFs, as well as
our multiple Gaussian component fitting technique, we plot the full
\ion{Ca}{II} near-IR triplet profile for SNe~2013bh ({\it top}) and
2000cx ({\it bottom}) in Figure~\ref{f:cair}. Both spectra were
obtained \about4~d after maximum and have had their linear
pseudo-continua (grey dotted line) removed. The data are the black
curves, the combined Gaussian fits are the grey solid curves, and the
individual Gaussian components are the grey short-dashed curves. The
separation between the HVFs (left-hand triplet of Gaussians) and the
PVFs (right-hand triplet of Gaussians) can easily be seen in both
objects. While we fully admit that the fits are not perfect and
fitting a set of Gaussians to the full profile is an
oversimplification, they do appear to capture the overall shapes of
both profiles, as well as at least some of the more subtle
details. The fits tend to have reduced $\chi^2$ values of 2--3. 

The vertical solid line is the \ion{Ca}{II} near-IR triplet velocity
for SN~2013bh, while the vertical long-dashed line is the
velocity for SN~2000cx. The PVFs in these two objects have
consistent expansion velocities at this epoch, however the HVFs of
SN~2013bh are found to be at larger velocities than those of SN~2000cx
(which can also be seen in Figure~\ref{f:lines}). Perhaps this
indicates a higher \ion{Ca}{II} optical depth in SN~2013bh. A more
detailed analysis of HVFs is beyond the scope of this paper, but an
in-depth study of HVFs in a large set of more normal SN~Ia spectra
will be undertaken in the near future (Silverman \etal in
preparation).

\subsection{Host Galaxy and Progenitor System}\label{ss:host}

SN~2013bh lies in the outskirts of the star-forming galaxy
SDSS~J150214.17+103843.6 at a redshift of $z = 0.0744$ \citep{Ahn12}, 
which, assuming a $\Lambda$CDM cosmology with $H_{0}=73$~\kms~\perMpc,
implies a distance of $326 \pm 22$~Mpc and a distance modulus of $\mu
= 37.57 \pm 0.15$~mag. It is located $15\farcs9$ west and $2\farcs0$
north of the centre of its host, a projected distance of
\about25.3~kpc \citep[or \about3 Petrosian
radii,][]{Petrosian76}. Correlation with a library of galaxy spectra
using the ``SuperNova IDentification'' code \citep[SNID;][]{Blondin07}
indicates that SDSS~J150214.17+103843.6 is a spiral (Sa/Sb) galaxy
with $z = 0.075$, quite close to the SDSS DR9 value of $z = 0.0744$
\citep{Ahn12}. While this is not an atypical type of host galaxy for a
SN~Ia \citep{Li11b}, very few (if any) SNe~Ia are found further away
from the centre of their host \citep[e.g.,][]{Wang13}.

While inner parts of the host galaxy of SN~2013bh may be currently
forming stars, the outer edges (where the SN is located) show no
obvious $u$-band emission \citep{Ahn12}, indicating a lack of star
formation \citep[$\la 4.9\times10^{-2}$~\msun~yr$^{-1}$;
e.g.,][]{Moustakas06}. In addition, typical metallicity gradients in
spiral galaxies \citep[e.g.,][]{Henry99} indicate that SN~2013bh went
off in a region that likely has relatively low-metallicity.

SN~2000cx also exploded in the periphery of its host galaxy, the
nearby (\about38~Mpc) S0 galaxy NGC~524. It was found 18~kpc (linear 
projected distance) from the centre of its host, in a low-metallicity
region \citep{Li01:00cx}. High-resolution spectra of SN~2000cx
presented by \citet{Patat07} do not show evidence of \ion{Na}{I}~D
absorption to very tight limits and they infer $N($\ion{Na}{I}$) \leq
2\times10^{10}$~cm$^{-2}$ and (for solar abundances) $N($H$) \leq
3\times10^{16}$~cm$^{-2}$. As mentioned above, our highest S/N
spectrum of SN~2013bh also did not show obvious absorption from
\ion{Na}{I}~D (with a 2$\sigma$ upper limit of 0.2~\AA) or evidence
for star formation from narrow emission lines.


Both SN~2013bh and SN~2000cx lack \ion{Na}{I}~D absorption which
indicates low host-galaxy reddening and a relatively ``clean''
circumstellar environment (i.e., free of expelled material from the
companion star). Their locations on the periphery of their host
galaxies, along with no obvious associated recent star formation,
point to old stellar population progenitors with relatively
low-metallicity for both objects. All of these observations are
consistent with many DD models \citep[e.g.,][]{Iben84}, though we
cannot definitively rule out all SD models \citep[e.g.,][]{Whelan73}.

\citet{Li01:00cx} and \citet{Candia03} found that none of the standard
SN~Ia light-curve fitting algorithms could match the observations of
SN~2000cx, and thus the same can be said for SN~2013bh; however, the
delayed-detonation DD3 model \citep{Woosley94,Pinto00:arxiv} did match
many of the observations of SN~2000cx. This model produced higher
kinetic energies and more nuclear burning than models of more normal
SNe~Ia (including the overluminous SN~1991T). As a result, the DD3
model produced a relatively large amount of $^{56}$Ni (\about1~\msun),
higher-than-normal blackbody temperatures, and high expansion
velocities. The latter two observables were seen in both SN~2000cx
\citep{Li01:00cx,Thomas04,Branch04:00cx} and SN~2013bh
(Sections~\ref{ss:synapps} and \ref{ss:vels}). Furthermore, this model
also predicted peak magnitudes and $\Delta m_{15}$ values consistent
with those observed for both SN~2000cx and SN~2013bh.

An alternative explanation for the HVFs observed in these two objects
comes from a 3D model presented by \citet{Thomas04}. They explain the
HVFs of \ion{Ca}{II} as coming from clumpy ejecta such that the
high-velocity material only partially covers the SN
photosphere. This idea is supported by \citet{Leonard00}, who found
significant polarization intrinsic SN~2000cx. \citet{Thomas04} also
derive (4.3--5.5)$\times 10^{-3}$~\msun\ for the high-velocity ejecta
mass in Ca, which they claim can be explained by primordial material
alone. This conclusion is consistent with the ``clean'' environment
determined for SNe~2000cx and 2013bh by the lack of \ion{Na}{I}~D or
narrow \ion{Ca}{II}~H\&K absorption from circumstellar
material. 


\section{Conclusions}\label{s:conclusions}

In this work we have presented optical and near-IR photometry of
SN~2013bh (aka iPTF13abc), a near twin of SN~2000cx, from
discovery through \about40~d past $r$-band maximum brightness and
low-resolution optical spectroscopy through \about20~d past
maximum. SN~2013dh reached a peak absolute magnitude of $M_r =
-19.2$~mag and had a decline rate of $\Delta m_{15}(r) = 0.73$~mag,
matching nearly identically to SN~2000cx. The colours of SN~2013bh
mostly follow those of SN~2000cx, but show evidence of a slightly
``flatter'' optical continuum that indicates a slightly lower
blackbody temperature than SN~2000cx. This is consistent with the 
relative spectral shapes of these two objects. The colours of
SNe~2000cx and 2013bh also indicate photospheric temperatures that are
higher than more normal SNe~Ia. This is supported by the presence of
strong absorption features from \ion{Fe}{III} and \ion{Si}{III} that
persist well past maximum brightness (which is not seen in any other
SN~Ia). Our spectral models from {\tt SYNAPPS} and direct measurements
of the strongest spectral features also indicate the presence of other 
iron-group elements (\ion{Co}{II}, \ion{Ni}{II}, \ion{Fe}{II}, and HVFs
of \ion{Ti}{II}) and intermediate-mass elements (\ion{Si}{II} and
\ion{S}{II}). Furthermore, in all of our spectra we observe separate
normal velocity features (\about12000~\kms) and HVFs
(\about24000~\kms) of \ion{Ca}{II}. 

The environments of both SN~2013bh and SN~2000cx appear to be
relatively ``clean,'' with negligible amounts of CSM, consistent with
their locations in the outskirts of their host galaxies and the lack
of obvious associated recent star formation. This points to the
progenitors of these objects being relatively low-metallicity, old
stars, consistent with the DD scenario (though the SD scenario cannot
be completely disproved). The delayed-detonation
DD3 model of \citet{Woosley94,Pinto00:arxiv} was successfully applied
to SN~2000cx and given the observational similarities with SN~2013bh,
it is reasonable to use it for this object as well. This model implies 
a relatively large amount of $^{56}$Ni produced (\about1~\msun),
higher-than-normal blackbody temperatures, and high expansion
velocities. The latter two of these were directly observed in
SN~2013bh. The model also predicts peak magnitudes and $\Delta m_{15}$
values consistent with those observed for both SN~2000cx and
SN~2013bh. A 3D model presented by \citet{Thomas04} explains the HVFs
of \ion{Ca}{II} as coming from (4.3--5.5)$\times 10^{-3}$~\msun\ of
clumpy ejecta, which can be explained by primordial material alone.

Of the \about2300 SNe~Ia given IAU designations from the beginning of
2000 through the end of 2012, SN~2000cx was the only one of its kind
discovered. PTF, which ran for 4 years starting in 2009, discovered
1250 SNe~Ia with no objects resembling SN~2000cx. iPTF, which began
in Jan~2013, discovered SN~2013bh (aka iPTF13abc) and through the end
of Jul~2013 has found 111 SNe~Ia. Very roughly, it seems that an
object like SN~2000cx or SN~2013bh is found in every \about2000
SNe~Ia. In other words, these objects are approximately 0.05\% of the
total SN~Ia rate. The extreme rarity of of these objects and how they
relate to more normal SNe~Ia and overluminous SN~1991T-like objects is
a challenge to SN progenitor models. As large-scale transient surveys 
(e.g., iPTF, Pan-STARRS, LSST) continue to (and will in the future)
find many new objects, more objects similar to SNe~2000cx and 2013bh
will likely be discovered.

\section*{Acknowledgments}
We would like to thank M.~Ganeshalingam, P.~Kelly, and E.~Ofek for
helpful discussions, J.~Caldwell, S.~Odewahn, and S.~Rostopchin for
their assistance with some of the observations, as well
as the PESSTO and CRTS collaborations for making some of their data on
SN~2013bh publicly available.
The HET is a joint project of the University of Texas at Austin, the
Pennsylvania State University, Stanford University,
Ludwig-Maximilians-Universit\"{a}t M\"{u}nchen, and
Georg-August-Universit\"{a}t G\"{o}ttingen. The HET is named in honor
of its principal benefactors, William P. Hobby and Robert
E. Eberly. The Marcario Low Resolution Spectrograph is named for Mike
Marcario of High Lonesome Optics who fabricated several optics for the
instrument but died before its completion. The LRS is a joint project
of the HET partnership and the Instituto de Astronom\'{i}a de la
Universidad Nacional Aut\'{o}noma de M\'{e}xico.
Some of the data presented herein were obtained at the W. M. Keck
Observatory, which is operated as a scientific partnership among the
California Institute of Technology, the University of California, and
the National Aeronautics and Space Administration (NASA); the
observatory was made possible by the generous financial support of the
W. M. Keck Foundation. The authors wish to recognize and acknowledge
the very significant cultural role and reverence that the summit of
Mauna Kea has always had within the indigenous Hawaiian community; we
are most fortunate to have the opportunity to conduct observations
from this mountain.
This work is partially based on observations made with the Nordic
Optical Telescope, operated on the island of La Palma jointly by
Denmark, Finland, Iceland, Norway, and Sweden, in the Spanish
Observatorio del Roque de los Muchachos of the Instituto de
Astrofisica de Canarias.
We thank the RATIR instrument team and the staff of the Observatorio
Astron\'omico Nacional on Sierra San Pedro M\'artir. RATIR is a
collaboration between the University of California, the Universidad
Nacional Auton\'oma de M\'exico, NASA Goddard Space Flight Center, and
Arizona State University, benefiting from the loan of an H2RG detector
from Teledyne Scientific and Imaging. RATIR, the automation of the
Harold L. Johnson Telescope of the Observatorio Astron\'omico Nacional
on Sierra San Pedro M\'artir, and the operation of both are funded by
the partner institutions and through NASA grants NNX09AH71G,
NNX09AT02G, NNX10AI27G, and NNX12AE66G, CONACyT grants
INFR-2009-01-122785, UNAM PAPIIT grant IN113810, and a UC
MEXUS-CONACyT grant.
The National Energy Research Scientific Computing Center, supported by
the Office of Science of the U.S. Department of Energy, provided
staff, computational resources, and data storage for this project. 
This research has made use of the NASA/IPAC Extragalactic Database
(NED) which is operated by the Jet Propulsion Laboratory, California
Institute of Technology, under contract with NASA.
Funding for SDSS-III has been provided by the Alfred P. Sloan
Foundation, the Participating Institutions, the NSF, and the
U.S. Department of Energy Office of Science. The SDSS-III web site is
http://www.sdss3.org/ .
JMS is supported by an NSF Astronomy and Astrophysics Postdoctoral
Fellowship under award AST-1302771.
JV is supported by Hungarian OTKA Grant NN 107637.
MMK acknowledges generous support from the Hubble Fellowship and
Carnegie-Princeton Fellowship.
JCW's supernova group at UT Austin is supported by NSF Grant AST
11-09801. Some work on this paper by JCW was done in the hospitable
clime of the Aspen Center for Physics that is supported by NSF Grant
PHY-1066293.
JSB acknowledges the generous support of a CDI grant (\#0941742) from
the National Science Foundation.

\bibliographystyle{mn2e}
\bibliography{/Users/jsilv/astro_refs}

\begin{thebibliography}{}

\bibitem[\protect\citeauthoryear{{Ahn} et~al.,}{{Ahn}  et~al.}{2012}]{Ahn12}
{Ahn} C.~P.,  et~al., 2012, \apjs, 203, 21

\bibitem[\protect\citeauthoryear{{Blondin} \& {Tonry}}{{Blondin} \&
  {Tonry}}{2007}]{Blondin07}
{Blondin} S.,  {Tonry} J.~L.,  2007, \apj, 666, 1024

\bibitem[\protect\citeauthoryear{{Bloom} et~al.,}{{Bloom}
  et~al.}{2012}]{Bloom12}
{Bloom} J.~S.,  et~al., 2012, \apjl, 744, L17

\bibitem[\protect\citeauthoryear{{Branch} et~al.,}{{Branch}
  et~al.}{2004}]{Branch04:00cx}
{Branch} D.,  et~al., 2004, \apj, 606, 413

\bibitem[\protect\citeauthoryear{{Butler} et~al.,}{{Butler}
  et~al.}{2012}]{Butler12}
{Butler} N.,  et~al., 2012, in Society of Photo-Optical Instrumentation
  Engineers (SPIE) Conference Series Vol.~8446 of Society of Photo-Optical
  Instrumentation Engineers (SPIE) Conference Series, {First Light with RATIR:
  An Automated 6-band Optical/NIR Imaging Camera}.
p.~34

\bibitem[\protect\citeauthoryear{{Buzzoni} et~al.,}{{Buzzoni}
  et~al.}{1984}]{Buzzoni84}
{Buzzoni} B.,  et~al., 1984, The Messenger, 38, 9

\bibitem[\protect\citeauthoryear{{Candia} et~al.,}{{Candia}
  et~al.}{2003}]{Candia03}
{Candia} P.,  et~al., 2003, \pasp, 115, 277

\bibitem[\protect\citeauthoryear{{Castor}}{{Castor}}{1970}]{Castor70}
{Castor} J.~I.,  1970, \mnras, 149, 111

\bibitem[\protect\citeauthoryear{{Cenko} et~al.,}{{Cenko}
  et~al.}{2006}]{Cenko06}
{Cenko} S.~B.,  et~al., 2006, \pasp, 118, 1396

\bibitem[\protect\citeauthoryear{{Childress} et~al.,}{{Childress}
  et~al.}{2013}]{Childress13:12fr}
{Childress} M.~J.,  et~al., 2013, \apj, 770, 29

\bibitem[\protect\citeauthoryear{{Childress}, {Filippenko}, {Ganeshalingam} \&
  {Schmidt}}{{Childress} et~al.}{2013}]{Childress13}
{Childress} M.~J.,  {Filippenko} A.~V.,  {Ganeshalingam} M.,    {Schmidt}
  B.~P.,  2013, \mnras, submitted (arXiv:1307.0563)

\bibitem[\protect\citeauthoryear{{Conley} et~al.,}{{Conley}
  et~al.}{2011}]{Conley11}
{Conley} A.,  et~al., 2011, \apjs, 192, 1

\bibitem[\protect\citeauthoryear{{Drake} et~al.,}{{Drake}
  et~al.}{2009}]{Drake09}
{Drake} A.~J.,  et~al., 2009, \apj, 696, 870

\bibitem[\protect\citeauthoryear{{Drake} et~al.,}{{Drake}
  et~al.}{2013}]{13bh:disc2}
{Drake} A.~J.,  et~al., 2013, Central Bureau Electronic Telegrams, 3480, 1

\bibitem[\protect\citeauthoryear{{Filippenko} et~al.,}{{Filippenko}
  et~al.}{1992}]{Filippenko92:91T}
{Filippenko} A.~V.,  et~al., 1992, \apjl, 384, L15

\bibitem[\protect\citeauthoryear{{Filippenko}, {Li}, {Treffers} \&
  {Modjaz}}{{Filippenko} et~al.}{2001}]{Filippenko01}
{Filippenko} A.~V.,  {Li} W.~D.,  {Treffers} R.~R.,    {Modjaz} M.,  2001, in
  {Paczy\'{n}ski} B.,  {Chen} W.~P.,   {Lemme} C.,  eds, {Small-Telescope
  Astronomy on Global Scales.} Vol.~246, {The Lick Observatory Supernova Search
  with the Katzman Automatic Imaging Telescope}.
Astron. Soc. Pac., San Francisco, p.~121

\bibitem[\protect\citeauthoryear{{Fisher}, {Branch}, {Nugent} \&
  {Baron}}{{Fisher} et~al.}{1997}]{Synow}
{Fisher} A.,  {Branch} D.,  {Nugent} P.,    {Baron} E.,  1997, \apjl, 481, L89

\bibitem[\protect\citeauthoryear{{Foley}}{{Foley}}{2012}]{Foley13:cahk}
{Foley} R.~J.,  2012, \mnras, submitted (arXiv:1212.6261)

\bibitem[\protect\citeauthoryear{{Foley} et~al.,}{{Foley}
  et~al.}{2003}]{Foley03}
{Foley} R.~J.,  et~al., 2003, \pasp, 115, 1220

\bibitem[\protect\citeauthoryear{{Fox} et~al.,}{{Fox}  et~al.}{2012}]{Fox12}
{Fox} O.~D.,  et~al., 2012, in Society of Photo-Optical Instrumentation
  Engineers (SPIE) Conference Series Vol.~8453 of Society of Photo-Optical
  Instrumentation Engineers (SPIE) Conference Series, {Performance and
  calibration of H2RG detectors and SIDECAR ASICs for the RATIR camera}.
p.~59

\bibitem[\protect\citeauthoryear{{Fox} \& {Filippenko}}{{Fox} \&
  {Filippenko}}{2013}]{Fox13}
{Fox} O.~D.,  {Filippenko} A.~V.,  2013, \apjl, in press (arXiv:1304.4934)

\bibitem[\protect\citeauthoryear{{Henry} \& {Worthey}}{{Henry} \&
  {Worthey}}{1999}]{Henry99}
{Henry} R.~B.~C.,  {Worthey} G.,  1999, \pasp, 111, 919

\bibitem[\protect\citeauthoryear{{Hill} et~al.,}{{Hill}  et~al.}{1998}]{Hill98}
{Hill} G.~J.,  et~al., 1998, in {D'Odorico} S.,  ed., Society of Photo-Optical
  Instrumentation Engineers (SPIE) Conference Series Vol.~3355 of Society of
  Photo-Optical Instrumentation Engineers (SPIE) Conference Series,
  {Hobby-Eberly Telescope low-resolution spectrograph: mechanical design}.
pp 433--443

\bibitem[\protect\citeauthoryear{{Howell}}{{Howell}}{2011}]{Howell11}
{Howell} D.~A.,  2011, Nature Communications, 2, 350

\bibitem[\protect\citeauthoryear{{Howell} et~al.,}{{Howell}
  et~al.}{2006}]{Howell06}
{Howell} D.~A.,  et~al., 2006, \nat, 443, 308

\bibitem[\protect\citeauthoryear{{Iben} Jr. \& {Tutukov}}{{Iben} \&
  {Tutukov}}{1984}]{Iben84}
{Iben} Jr. I.,  {Tutukov} A.~V.,  1984, \apjs, 54, 335

\bibitem[\protect\citeauthoryear{{Jeffery}}{{Jeffery}}{1989}]{Jeffery89}
{Jeffery} D.~J.,  1989, \apjs, 71, 951

\bibitem[\protect\citeauthoryear{{Jordi}, {Grebel} \& {Ammon}}{{Jordi}
  et~al.}{2006}]{Jordi06}
{Jordi} K.,  {Grebel} E.~K.,    {Ammon} K.,  2006, \aap, 460, 339

\bibitem[\protect\citeauthoryear{{Kasliwal} et~al.,}{{Kasliwal}
  et~al.}{2012}]{Kasliwal12}
{Kasliwal} M.~M.,  et~al., 2012, \apj, 755, 161

\bibitem[\protect\citeauthoryear{{Kulkarni}}{{Kulkarni}}{2013}]{Kulkarni13}
{Kulkarni} S.~R.,  2013, The Astronomer's Telegram, 4807, 1

\bibitem[\protect\citeauthoryear{{Law} et~al.,}{{Law}  et~al.}{2009}]{Law09}
{Law} N.~M.,  et~al., 2009, \pasp, 121, 1395

\bibitem[\protect\citeauthoryear{{Leonard}, {Filippenko}, {Chornock} \&
  {Li}}{{Leonard} et~al.}{2000}]{Leonard00}
{Leonard} D.~C.,  {Filippenko} A.~V.,  {Chornock} R.,    {Li} W.~D.,  2000,
  \iaucirc, 7471, 3

\bibitem[\protect\citeauthoryear{{Li} et~al.,}{{Li}  et~al.}{2001}]{Li01:00cx}
{Li} W.,  et~al., 2001, \pasp, 113, 1178

\bibitem[\protect\citeauthoryear{{Li} et~al.,}{{Li}  et~al.}{2011}]{Li11b}
{Li} W.,  et~al., 2011, \mnras, 412, 1473

\bibitem[\protect\citeauthoryear{{Lira} et~al.,}{{Lira}  et~al.}{1998}]{Lira98}
{Lira} P.,  et~al., 1998, \aj, 115, 234

\bibitem[\protect\citeauthoryear{{Marion} et~al.,}{{Marion}
  et~al.}{2013}]{Marion13:09ig}
{Marion} G.~H.,  et~al., 2013, \apj, submitted (arXiv:1302.3537)

\bibitem[\protect\citeauthoryear{{Matheson} et~al.,}{{Matheson}
  et~al.}{2008}]{Matheson08}
{Matheson} T.,  et~al., 2008, \aj, 135, 1598

\bibitem[\protect\citeauthoryear{{Matheson}, {Filippenko}, {Ho}, {Barth} \&
  {Leonard}}{{Matheson} et~al.}{2000}]{Matheson00:93j}
{Matheson} T.,  {Filippenko} A.~V.,  {Ho} L.~C.,  {Barth} A.~J.,    {Leonard}
  D.~C.,  2000, \aj, 120, 1499

\bibitem[\protect\citeauthoryear{{Maund} et~al.,}{{Maund}
  et~al.}{2013}]{Maund12}
{Maund} J.~R.,  et~al., 2013, \mnras, 431, L102

\bibitem[\protect\citeauthoryear{{Mazzali} et~al.,}{{Mazzali}
  et~al.}{2005}]{Mazzali05}
{Mazzali} P.~A.,  et~al., 2005, \apjl, 623, L37

\bibitem[\protect\citeauthoryear{{Morales-Garoffolo}
  et~al.,}{{Morales-Garoffolo}  et~al.}{2013}]{13bh:disc}
{Morales-Garoffolo} A.,  et~al., 2013, The Astronomer's Telegram, 4955, 1

\bibitem[\protect\citeauthoryear{{Nugent} et~al.,}{{Nugent}
  et~al.}{2011}]{Nugent11}
{Nugent} P.~E.,  et~al., 2011, \nat, 480, 344

\bibitem[\protect\citeauthoryear{{Oke}, {Cohen}, {Carr} et~al.,}{{Oke}
  et~al.}{1995}]{Oke95}
{Oke} J.~B.,  {Cohen} J.~G.,  {Carr} M.,    et~al., 1995, \pasp, 107, 375

\bibitem[\protect\citeauthoryear{{Oke} \& {Gunn}}{{Oke} \&
  {Gunn}}{1982}]{Oke82}
{Oke} J.~B.,  {Gunn} J.~E.,  1982, \pasp, 94, 586

\bibitem[\protect\citeauthoryear{{Parrent} et~al.,}{{Parrent}
  et~al.}{2012}]{Parrent12}
{Parrent} J.~T.,  et~al., 2012, \apjl, 752, L26

\bibitem[\protect\citeauthoryear{{Patat} et~al.,}{{Patat}
  et~al.}{2007}]{Patat07}
{Patat} F.,  et~al., 2007, Science, 317, 924

\bibitem[\protect\citeauthoryear{{Perlmutter} et~al.,}{{Perlmutter}
  et~al.}{1999}]{Perlmutter99}
{Perlmutter} S.,  et~al., 1999, \apj, 517, 565

\bibitem[\protect\citeauthoryear{{Petrosian}}{{Petrosian}}{1976}]{Petrosian76}
{Petrosian} V.,  1976, \apjl, 209, L1

\bibitem[\protect\citeauthoryear{{Phillips}}{{Phillips}}{1993}]{Phillips93}
{Phillips} M.~M.,  1993, \apjl, 413, L105

\bibitem[\protect\citeauthoryear{{Phillips}, {Wells}, {Suntzeff}, {Hamuy},
  {Leibundgut}, {Kirshner} \& {Foltz}}{{Phillips} et~al.}{1992}]{Phillips92}
{Phillips} M.~M.,  {Wells} L.~A.,  {Suntzeff} N.~B.,  {Hamuy} M.,  {Leibundgut}
  B.,  {Kirshner} R.~P.,    {Foltz} C.~B.,  1992, \aj, 103, 1632

\bibitem[\protect\citeauthoryear{{Pinto} \& {Eastman}}{{Pinto} \&
  {Eastman}}{2000a}]{Pinto00}
{Pinto} P.~A.,  {Eastman} R.~G.,  2000a, \apj, 530, 757

\bibitem[\protect\citeauthoryear{{Pinto} \& {Eastman}}{{Pinto} \&
  {Eastman}}{2000b}]{Pinto00:arxiv}
{Pinto} P.~A.,  {Eastman} R.~G.,  2000b, \apj, submitted (arXiv:000.6171)

\bibitem[\protect\citeauthoryear{{Poznanski}, {Ganeshalingam}, {Silverman} \&
  {Filippenko}}{{Poznanski} et~al.}{2011}]{Poznanski11}
{Poznanski} D.,  {Ganeshalingam} M.,  {Silverman} J.~M.,    {Filippenko} A.~V.,
   2011, \mnras, 415, L81

\bibitem[\protect\citeauthoryear{{Rau} et~al.,}{{Rau}  et~al.}{2009}]{Rau09}
{Rau} A.,  et~al., 2009, \pasp, 121, 1334

\bibitem[\protect\citeauthoryear{{Richmond}, {Treffers} \&
  {Filippenko}}{{Richmond} et~al.}{1993}]{Richmond93}
{Richmond} M.,  {Treffers} R.~R.,    {Filippenko} A.~V.,  1993, \pasp, 105,
  1164

\bibitem[\protect\citeauthoryear{{Riess} et~al.,}{{Riess}
  et~al.}{1998}]{Riess98:lambda}
{Riess} A.~G.,  et~al., 1998, \aj, 116, 1009

\bibitem[\protect\citeauthoryear{{Rudy}, {Lynch}, {Mazuk}, {Venturini},
  {Puetter} \& {H{\"o}flich}}{{Rudy} et~al.}{2002}]{Rudy02}
{Rudy} R.~J.,  {Lynch} D.~K.,  {Mazuk} S.,  {Venturini} C.~C.,  {Puetter}
  R.~C.,    {H{\"o}flich} P.,  2002, \apj, 565, 413

\bibitem[\protect\citeauthoryear{{Scalzo} et~al.,}{{Scalzo}
  et~al.}{2010}]{Scalzo10}
{Scalzo} R.~A.,  et~al., 2010, \apj, 713, 1073

\bibitem[\protect\citeauthoryear{{Schlegel}, {Finkbeiner} \&
  {Davis}}{{Schlegel} et~al.}{1998}]{Schlegel98}
{Schlegel} D.~J.,  {Finkbeiner} D.~P.,    {Davis} M.,  1998, \apj, 500, 525

\bibitem[\protect\citeauthoryear{{Silverman} et~al.,}{{Silverman}
  et~al.}{2012a}]{Silverman12:BSNIPI}
{Silverman} J.~M.,  et~al., 2012a, \mnras, 425, 1789

\bibitem[\protect\citeauthoryear{{Silverman} et~al.,}{{Silverman}
  et~al.}{2012b}]{Silverman12:12cg}
{Silverman} J.~M.,  et~al., 2012b, \apjl, 756, L7

\bibitem[\protect\citeauthoryear{{Silverman} et~al.,}{{Silverman}
  et~al.}{2013}]{Silverman13:iacsm}
{Silverman} J.~M.,  et~al., 2013, \apjs, accepted (arXiv:1303.0763)

\bibitem[\protect\citeauthoryear{{Silverman}, {Ganeshalingam}, {Li},
  {Filippenko}, {Miller} \& {Poznanski}}{{Silverman}
  et~al.}{2011}]{Silverman11}
{Silverman} J.~M.,  {Ganeshalingam} M.,  {Li} W.,  {Filippenko} A.~V.,
  {Miller} A.~A.,    {Poznanski} D.,  2011, \mnras, 410, 585

\bibitem[\protect\citeauthoryear{{Silverman}, {Kong} \&
  {Filippenko}}{{Silverman} et~al.}{2012}]{Silverman12:BSNIPII}
{Silverman} J.~M.,  {Kong} J.~J.,    {Filippenko} A.~V.,  2012, \mnras, 425,
  1819

\bibitem[\protect\citeauthoryear{{Skrutskie} et~al.,}{{Skrutskie}
  et~al.}{2006}]{Skrutskie06}
{Skrutskie} M.~F.,  et~al., 2006, \aj, 131, 1163

\bibitem[\protect\citeauthoryear{{Sobolev}}{{Sobolev}}{1960}]{Sobolev60}
{Sobolev} V.~V.,  1960, {Moving envelopes of stars}.
Cambridge: Harvard University Press

\bibitem[\protect\citeauthoryear{{Sollerman} et~al.,}{{Sollerman}
  et~al.}{2004}]{Sollerman04}
{Sollerman} J.,  et~al., 2004, \aap, 428, 555

\bibitem[\protect\citeauthoryear{{Sullivan} et~al.,}{{Sullivan}
  et~al.}{2011}]{Sullivan11}
{Sullivan} M.,  et~al., 2011, \apj, 737, 102

\bibitem[\protect\citeauthoryear{{Suzuki} et~al.,}{{Suzuki}
  et~al.}{2012}]{Suzuki12}
{Suzuki} N.,  et~al., 2012, \apj, 746, 85

\bibitem[\protect\citeauthoryear{{Thomas}, {Branch}, {Baron}, {Nomoto}, {Li} \&
  {Filippenko}}{{Thomas} et~al.}{2004}]{Thomas04}
{Thomas} R.~C.,  {Branch} D.,  {Baron} E.,  {Nomoto} K.,  {Li} W.,
  {Filippenko} A.~V.,  2004, \apj, 601, 1019

\bibitem[\protect\citeauthoryear{{Thomas}, {Nugent} \& {Meza}}{{Thomas}
  et~al.}{2011}]{Thomas11:synapps}
{Thomas} R.~C.,  {Nugent} P.~E.,    {Meza} J.~C.,  2011, \pasp, 123, 237

\bibitem[\protect\citeauthoryear{{Valenti} et~al.,}{{Valenti}
  et~al.}{2011}]{Valenti11}
{Valenti} S.,  et~al., 2011, \mnras, 416, 3138

\bibitem[\protect\citeauthoryear{{Vink{\'o}} et~al.,}{{Vink{\'o}}
  et~al.}{2012}]{Vinko12}
{Vink{\'o}} J.,  et~al., 2012, \aap, 546, A12

\bibitem[\protect\citeauthoryear{{Wade} \& {Horne}}{{Wade} \&
  {Horne}}{1988}]{Wade88}
{Wade} R.~A.,  {Horne} K.,  1988, \apj, 324, 411

\bibitem[\protect\citeauthoryear{{Wang}, {Wang}, {Filippenko}, {Zhang} \&
  {Zhao}}{{Wang} et~al.}{2013}]{Wang13}
{Wang} X.,  {Wang} L.,  {Filippenko} A.~V.,  {Zhang} T.,    {Zhao} X.,  2013,
  Science, 340, 170

\bibitem[\protect\citeauthoryear{{Watson} et~al.,}{{Watson}
  et~al.}{2012}]{Watson12}
{Watson} A.~M.,  et~al., 2012, in Society of Photo-Optical Instrumentation
  Engineers (SPIE) Conference Series Vol.~8444 of Society of Photo-Optical
  Instrumentation Engineers (SPIE) Conference Series, {Automation of the
  OAN/SPM 1.5-meter Johnson telescope for operations with RATIR}

\bibitem[\protect\citeauthoryear{{Webbink}}{{Webbink}}{1984}]{Webbink84}
{Webbink} R.~F.,  1984, \apj, 277, 355

\bibitem[\protect\citeauthoryear{{Whelan} \& {Iben} Jr.}{{Whelan} \&
  {Iben}}{1973}]{Whelan73}
{Whelan} J.,  {Iben} Jr. I.,  1973, \apj, 186, 1007

\bibitem[\protect\citeauthoryear{{Woosley} \& {Weaver}}{{Woosley} \&
  {Weaver}}{1994}]{Woosley94}
{Woosley} S.~E.,  {Weaver} T.~A.,  1994, \apj, 423, 371

\bibitem[\protect\citeauthoryear{{Yaron} \& {Gal-Yam}}{{Yaron} \&
  {Gal-Yam}}{2012}]{Yaron12}
{Yaron} O.,  {Gal-Yam} A.,  2012, \pasp, 124, 668

\bibitem[\protect\citeauthoryear{{Yu}, {Modjaz} \& {Li}}{{Yu}
  et~al.}{2000}]{00cx:disc}
{Yu} C.,  {Modjaz} M.,    {Li} W.~D.,  2000, \iaucirc, 7458, 1

\end{thebibliography}

\label{lastpage}

\end{document}